\documentclass[12pt, reqno]{amsart}

\usepackage[a4paper, margin=1.in, foot=.3in]{geometry}

\usepackage{color}

\usepackage{amsfonts,amssymb,amsmath,amscd}

\usepackage{pstricks}

\definecolor{darkblue}{rgb}{0,0,.8}
\definecolor{darkred}{rgb}{.8,0,0}
\definecolor{darkgreen}{rgb}{0,0.8,0}

\usepackage{hyperref}
\hypersetup{
    pdfstartview=FitH,
    colorlinks,
    citecolor=darkgreen,
    linkcolor=darkred,
    urlcolor=darkblue
}

\usepackage[noBBpl,slantedGreek]{mathpazo}

\allowdisplaybreaks \numberwithin{equation}{section}

\makeatletter
\def\subsection{\@startsection{subsection}{2}%
  \z@{.5\linespacing\@plus.7\linespacing}{-.5em}%
  {\normalfont\bfseries\mathversion{bold}}}
\makeatother

\psset{unit=1pt, dimen=middle, linewidth=.5pt, arrowsize=3pt 2}

\newdimen \psx
\newdimen \psy

\def\psddots (#1,#2){
  \psx #1\psxunit
  \psy #2\psyunit
  \qdisk(\psx,\psy){.7pt}
  \advance \psx -.2\psxunit
  \advance \psy .2\psyunit
  \qdisk(\psx,\psy){.7pt}
  \advance \psx .4\psxunit
  \advance \psy -.4\psyunit
  \qdisk(\psx,\psy){.7pt}
}

\def\pssddots (#1,#2){
  \psx #1\psxunit
  \psy #2\psyunit
  \qdisk(\psx,\psy){.7pt}
  \advance \psx .2\psxunit
  \advance \psy .2\psyunit
  \qdisk(\psx,\psy){.7pt}
  \advance \psx -.4\psxunit
  \advance \psy -.4\psyunit
  \qdisk(\psx,\psy){.7pt}
}

\def \bm #1{\mbox{\boldmath $#1$}}

\def \qbinom #1#2{\begin{bmatrix} #1 \\ #2 \end{bmatrix}}

\def \wt {\widetilde}

\def \bbC {\mathbb C}

\def \bbZ {\mathbb Z}

\def \calI {\mathcal I}
\def \calK {\mathcal K}

\def \calN {\mathcal N}
\def \calR {\mathcal R}

\def \gothb {\mathfrak b}
\def \gothg {\mathfrak g}
\def \gothh {\mathfrak h}

\def \gothsl {\mathfrak{sl}}
\def \rmOsc {\mathrm{Osc}}
\def \rmd {\mathrm d}
\def \rme {\mathrm e}

\def \End {\mathrm{End}}

\begin{document}

\title[On the universal $R$-matrix for the Izergin--Korepin model]{On the universal
$\bm R$-matrix for the Izergin--Korepin model}

\author[H. Boos]{Herman Boos}
\address{Fachbereich C -- Physik, Bergische Universit\"at Wuppertal, 42097 Wuppertal, Germany}
\email{boos@physik.uni-wuppertal.de}

\author[F. G\"ohmann]{Frank G\"ohmann}
\address{Fachbereich C -- Physik, Bergische Universit\"at Wuppertal, 42097 Wuppertal, Germany}
\email{goehmann@physik.uni-wuppertal.de}

\author[A. Kl\"umper]{Andreas Kl\"umper}
\address{Fachbereich C -- Physik, Bergische Universit\"at Wuppertal, 42097 Wuppertal, Germany}
\email{kluemper@uni-wuppertal.de}

\author[Kh. S. Nirov]{\vskip .2em Khazret S. Nirov}
\address{Institute for Nuclear Research of the Russian Academy of Sciences, 60th October Ave 7a,
117312 Moscow, Russia} \curraddr{Fachbereich C -- Physik, Bergische Universit\"at Wuppertal, 42097
Wuppertal, Germany} \email{knirov@physik.uni-wuppertal.de}

\author[A. V. Razumov]{Alexander V. Razumov}
\address{Institute for High Energy Physics, 142281 Protvino, Moscow region, Russia}
\email{Alexander.Razumov@ihep.ru}

\begin{abstract}
We continue our exercises with the universal $R$-matrix based on the Kho\-rosh\-kin and Tols\-toy
formula. Here we present our results for the case of the twisted affine Kac--Moody Lie algebra of
type $A^{(2)}_2$. Our interest in this case is inspired by the fact that the Tzitz\'eica equation
is associated with $A^{(2)}_2$ in a similar way as the sine-Gordon equation is related to
$A^{(1)}_1$. The fundamental spin-chain Hamiltonian is constructed systematically as the
logarithmic derivative of the transfer matrix. $L$-operators of two types are obtained by using
$q$-deformed oscillators.
\end{abstract}


\maketitle

\tableofcontents

\section{Introduction}
\label{int}

The Izergin--Korepin model \cite{IzeKor81} was introduced as a quantum integrable model related to
a classical integrable system described by the nonlinear differential equation
\[
\partial_+ \partial_- F = -m^2 [\exp(-2F) - \exp(F)]
\]
for a function $F$ of two independent variables. Although it was introduced for the first time by
Tzitz\'eica within the framework of differential geometry, this equation is mostly known as the
Dodd--Bullough--Mikhailov or Jiber--Mikhailov--Shabat model, since it was investigated later in the
context of the theory of solitons and general aspects of classical integrability. It is also known
that this equation is a particular case of the Toda equations associated with twisted loop groups
\cite{NirRaz08a, NirRaz08b}. Here the affine group of type $A^{(2)}_2$ plays the role of the
underlying symmetry group, which for the sine-Gordon equation takes the simplest affine group of
type $A^{(1)}_1$.

To a certain extent, the model under consideration has features reminiscent of both, the so-called
$\gothsl_3$ and $\gothsl_2$, cases: on one hand it is related to a three-dimensional matrix
representation, while, on the other hand, the infinite-dimensional part of the whole representation
should correspond to only one scalar field entering the integrable equation written above, and
therefore we expect that a single copy of the $q$-deformed oscillator algebra will be sufficient to
describe $L$- and $Q$-operators.

We work in the spirit and use notations of \cite{BooGoeKluNirRaz10}, continuing our exercises with
the universal $R$-matrix based on the remarkable formula presented by Khoroshkin and Tolstoy
\cite{KhoTol91, KhoTol92, TolKho92}. After reproducing the $R$-matrix of the Izergin--Korepin
model, we construct two types of the $L$-operators using a $q$-deformed oscillator algebra. In the
case under consideration these $L$-operators turn out to be related by a similarity transformation.
A similar relation holds for the $R$-matrix as well. We also discuss various (anti)automorphisms
generating new $L$-operators from given ones. The $L$-operators can then be used for the subsequent
construction of $Q$-operators encoding the information about the spectrum of the quantum integrable
system under consideration.

The problem of the calculation of the eigenvalues of the transfer-matrix for the Izergin--Korepin
model was studied in \cite{Res83}. An algebraic Bethe ansatz solution of this model was constructed
in \cite{Res87, Tar88}. This consideration was extended in \cite{MezNep92, LiShiYue03} to the case
of open boundary conditions.

\section{Generators and roots}
\label{rig}

Let $A = (a_{ij})_{i, j = 0, 1}$ be the generalized Cartan matrix of type $A_2^{(2)}$ having the explicit form
\[
A = \left(\begin{array}{rr} 2 & -1 \\
                           -4 & 2
           \end{array}\right).
\]
The matrix $A$ is symmetrizable and we have $d_i a_{ij} = d_j a_{ji}$ for $d_0 = 2$ and $d_1 =
1/2$. We denote
\[
A^S = (a_{ij}^S) = (d_i a_{ij}) = \left(\begin{array}{rr} 4 & -2 \\
                           -2 & 1
           \end{array}\right).
\]

Before proceeding to the universal $R$-matrix, let us first describe the structure of the
Kac--Moody algebra $\gothg'(A)$, enveloping algebra $U(\gothg'(A))$ and its quantum deformation
$U_\hbar(\gothg'(A))$. In accordance with the general construction, see, for example, the
books~\cite{Kac94, Car05}, the twisted affine Lie algebra $\gothg'(A)$ is generated by the elements
$h_i$, $e_i$, $f_i$, $i = 0, 1$, with the defining relations
\begin{gather}
[h_i,h_j] = 0, \label{dr1} \\
[h_i,e_j] = a_{ij} e_j, \qquad [h_i,f_j] = -a_{ij} f_j, \label{dr2} \\
[e_i,f_j] = \delta_{ij} h_i, \label{dr3}
\end{gather}
satisfied for all $i$ and $j$, and the Serre relations
\begin{equation}
(\mathrm{ad} \, e_i)^{1-{a_{ij}}}(e_j) = 0, \qquad (\mathrm{ad} \, f_i)^{1-{a_{ij}}}(f_j) = 0, \label{sr}
\end{equation}
satisfied for all distinct $i$ and $j$.

We denote the linear span of the generators $h_i$ by $\gothh'(A)$ and its dual space by
$\gothh'^*(A)$. The vector space $\gothh'(A)$ is the Cartan subalgebra of $\gothg'(A)$, and we have
the decomposition
\[
\gothg'(A) = \gothh'(A) \oplus \bigoplus_{\gamma \in \triangle(A)} \gothg'(A)_\gamma,
\]
where for any $\gamma \in \gothh'^*(A)$ we denote
\[
\gothg'(A)_\gamma = \{ x \in \gothg'(A) \mid [h, x] = \gamma(x) x \mbox{ for any } h \in \gothh'(A) \}
\]
and
\[
\triangle(A) = \{ \gamma \in \gothh'^*(A) \mid \gamma \ne 0, \, \gothg'(A)_\gamma \ne \{0\} \}.
\]
The elements of $\triangle(A)$ are roots, and $\gothg'(A)_\gamma$ is the root space of $\gamma$ whose nonzero
elements are root vectors.

The generators $e_i$ are evidently root vectors. We denote the corresponding roots by $\alpha_i$. These are
simple roots.  Any other root is a linear combination of simple roots with integer coefficients all of which
are either non-negative or non-positive. In the former case we have a positive root and in the latter a negative
one. In particular, the generators $f_i$ are root vectors corresponding to the negative roots $-\alpha_i$. One
has $\triangle(A) = \triangle_+(A) \sqcup \triangle_-(A)$, where $\triangle_+(A)$ is the set of positive roots
and $\triangle_-(A) = - \triangle_+(A)$.

The symmetrized Cartan matrix $A^S$ determines a symmetric bilinear form on $\gothh'^*(A)$ by the equality
\[
(\alpha_i, \alpha_j) = a_{ij}^S.
\]
Explicitly we have
\[
(\alpha_0,\alpha_0) = 4, \qquad (\alpha_0,\alpha_1) = -2, \qquad (\alpha_1,\alpha_0) = -2,
\qquad (\alpha_1,\alpha_1) = 1.
\]
It is convenient to denote\footnote{In fact, $\delta$ is the minimal positive imaginary root \cite{Kac94, Car05}.}
\[
\delta = \alpha_0 + 2 \alpha_1, \qquad \alpha = \alpha_1,
\]
so that
\[
(\delta, \delta) = (\delta, \alpha) = (\alpha,\delta) = 0, \qquad (\alpha, \alpha) = 1.
\]

It can be shown that the set of the positive roots is
\begin{multline*}
\triangle_+(A) = \{ \alpha + m \delta \mid m \in \bbZ_{\ge 0} \} \cup \{ 2 \alpha +
(2 m + 1) \delta \mid m \in \bbZ_{\ge 0} \} \\
\cup \{m \delta \mid m \in \bbZ_{>0} \} \cup \{ \delta - 2 \alpha + 2 m \delta \mid m \in \bbZ_{\ge 0} \}
\cup \{ \delta - \alpha + m \delta \mid m \in \bbZ_{\ge 0} \},
\end{multline*}
see, for example, the book \cite{Car05}. All root spaces corresponding to positive and negative roots
are one-dimensional. Choosing a root vector for each root and adding the Cartan generators $h_i$ we obtain
a Cartan basis of $\gothg'(A)$.

The enveloping algebra $U(\gothg'(A))$ is defined as the unital associative algebra with generators
$h_i$, $e_i$, $f_i$ and with the same relations (\ref{dr1})--(\ref{sr}) as $\gothg'(A)$. Here we can
rewrite the Serre relations (\ref{sr}) as
\begin{gather}
\sum_{k = 0}^{1 - a_{ij}} (-1)^k \binom{1 - a_{ij}}{k} (e_i)^{1 - a_{ij} - k} \,
e_j \, (e_i)^k = 0, \label{usr1} \\
\sum_{k = 0}^{1 - a_{ij}} (-1)^k \binom{1 - a_{ij}}{k} (f_i)^{1 - a_{ij} - k} \,
f_j \, (f_i)^k = 0. \label{usr2}
\end{gather}
The Lie algebra $\gothg'(A)$ can be naturally considered as a subspace of $U(\gothg'(A))$, and a
Cartan basis of $\gothg'(A)$ as any of its bases generates a Poincar\'e--Birkhoff--Witt basis of
$U(\gothg'(A))$.

Let $\hbar$ be an indeterminate and $q = \exp \hbar$. In accordance with the general definition,
see, for example, the book \cite{ChaPre95}, the quantum deformation of $U(\gothg'(A))$, the quantum
group $U_\hbar(\gothg'(A))$, is the topological $\bbC[[\hbar]]$-algebra generated by six elements
$h_i$, $e_i$, $f_i$, $i = 0, 1$, with the defining relations
\begin{gather}
[h_i, h_j] = 0, \label{aqdr1}\\
[h_i, e_j] = a_{ij} e_j, \qquad
[h_i, f_j] = -a_{ij} f_j, \label{aqdr2}\\
[e_i, f_j] = \delta_{ij} \frac{q^{d_i h_i} -
q^{-d_i h_i}}{q^{d_i} - q^{-d_i}}, \label{aqdr3}
\end{gather}
satisfied for all $i$ and $j$, and the Serre relations
\begin{gather}
\sum_{k = 0}^{1 - a_{ij}} (-1)^k \qbinom{1 - a_{ij}}{k}_{q^{d_i}} (e_i)^{1 - a_{ij} - k} \,
e_j \, (e_i)^k = 0, \label{q-serre1}\\
\sum_{k = 0}^{1 - a_{ij}} (-1)^k \qbinom{1 - a_{ij}}{k}_{q^{d_i}} (f_i)^{1 - a_{ij} - k} \,
f_j \, (f_i)^k = 0, \label{q-serre2}
\end{gather}
satisfied for all distinct $i$ and $j$. Here we have introduced the $q$-binomial coefficients
\[
\qbinom{n}{m}_q = \frac{[n]_q!}{[n-m]_q! [m]_q!}, \qquad [n]_q! = [n]_q [n-1]_q \cdots [1]_q,
\qquad [n]_q = \frac{q^n - q^{-n}}{q-q^{-1}}.
\]
It is clear that the relations (\ref{aqdr1})--(\ref{aqdr3}) are deformations of the relations
(\ref{dr1})--(\ref{dr3}), and the quantum Serre relations (\ref{q-serre1}), (\ref{q-serre2})
are deformations of the Serre relations (\ref{usr1}),~(\ref{usr2}).
Khoroshkin and Tolstoy use a slightly different definition of a quantum group \cite{TolKho92, KhoTol92}.
We come to this definition using the rescaling of the generators
\[
h_i \to d_i^{-1} h_i^{\mathstrut}, \qquad
e_i \to \left( \frac{q - q^{-1}}{q^{d_i} - q^{-d_i}} \right)^{1/2} e_i,
\qquad f_i \to \left( \frac{q - q^{-1}}{q^{d_i} - q^{-d_i}} \right)^{1/2} f_i.
\]
After this, the defining relations (\ref{aqdr1})--(\ref{aqdr3}) take the form
\begin{gather}
[h_i, h_j] = 0, \label{qdr1}\\
[h_i, e_j] = a^S_{ij} e_j^{\mathstrut}, \qquad
[h_i, f_j] = -a^S_{ij} f_j^{\mathstrut}, \label{qdr2}\\
[e_i, f_j] = \delta_{ij} \frac{q^{h_i} -
q^{-h_i}}{q - q^{-1}}, \label{qdr3}
\end{gather}
while the Serre relations (\ref{q-serre1}) and (\ref{q-serre2}) preserve their form. Below we work in terms
of the rescaled generators. Note that the element $c = h_0 + 2 h_1$ belongs to the center of $U_\hbar(\gothg'(A))$.
It is convenient to assume that the definition of $U_\hbar(\gothg'(A))$ includes an additional relation
\[
c = h_0 + 2 h_1 = 0.
\]
This allows us to use for the case of the quantum groups $U_\hbar(\gothg'(A))$ the
Khoroshkin--Tolstoy formula for the universal $R$-matrix valid for quantum groups
$U_\hbar(\gothg(A))$ just putting there $c = 0$.

The quantum group $U_\hbar(\gothg'(A))$ is a topological Hopf $\bbC[[\hbar]]$-algebra with correspondingly
defined comultiplication, counit and antipode, see, for example, the book \cite{ChaPre95}.

Associate now with each root from the infinite root system $\triangle(A)$ a corresponding root
vector. We denote the root vector corresponding to a positive root $\gamma$ by $e_\gamma$, and the
root vector corresponding to a negative root $-\gamma$ by $f_\gamma$. Let us start with the simple
positive roots and put
\[
e_{\alpha_0} = e_{\delta - 2 \alpha} = e_0, \qquad e_{\alpha_1} = e_\alpha = e_1.
\]
The higher root vectors corresponding to positive roots are defined recursively by the
relations\footnote{We use primed notation for the root vectors corresponding to the roots
$m \delta$ because we redefine them below.}
\begin{gather}
e_{\delta-\alpha} = ([2]_q)^{-1/2} [e_\alpha,e_{\delta-2\alpha}]_q, \qquad e'_\delta =
[e_\alpha,e_{\delta-\alpha}]_q,  \label{hivec1} \\
e_{\alpha+m\delta} = ([3]_{q^{1/2}})^{-1} [e_{\alpha+(m-1)\delta},e'_\delta]_q, \qquad
e_{\delta-\alpha+m\delta} = ([3]_{q^{1/2}})^{-1} [e'_\delta,e_{\delta-\alpha+(m-1)\delta)}]_q,
\label{hivec2} \\
e_{2\alpha+(2m+1)\delta} = ([2]_q)^{-1/2} [e_{\alpha+m\delta},e_{\alpha+(m+1)\delta}]_q, \label{hivec3}\\
e_{\delta-2\alpha+2(m+1)\delta} = ([2]_q)^{-1/2}
[e_{\delta-\alpha+(m+1)\delta},e_{\delta-\alpha+m\delta}]_q, \label{hivec4} \\
e'_{m\delta} = [e_{\alpha+(m-1)\delta},e_{\delta-\alpha}]_q, \label{hivec5}
\end{gather}
where the $q$-deformed commutator is defined as
\[
[e_\gamma, e_{\gamma'}]_q = e_\gamma e_{\gamma'} - q^{(\gamma, \gamma')} e_{\gamma'} e_\gamma
\]
for any two roots $\gamma$ and $\gamma'$ from the root system $\triangle_+(A)$. These relations allow us
to construct the root vectors corresponding to all the roots from the set $\triangle_+(A)$. In fact, the
expression for the universal $R$-matrix given by Khoroshkin and Tolstoy \cite{TolKho92} contains the root
vectors $e_{m \delta}$ related to the root vectors $e'_{m \delta}$ by the equality
\[
(q-q^{-1}) e_\delta(x) = \log[1 + (q - q^{-1})e^{\prime}_\delta(x)],
\]
where
\[
e^{\prime}_\delta(x) = \sum_{m>0} e^{\prime}_{m\delta} x^{-m}, \qquad e_\delta(x) = \sum_{m>0}
e_{m\delta} x^{-m}.
\]

To complete the set of root vectors we need to construct root vectors for negative roots from the system
$\triangle_-(A)$. We do it using the rule
\[
f_\gamma = \omega(e_{\gamma}) = e_{-\gamma}
\]
for any $\gamma \in \triangle_+(A)$. Here $\omega$ is the Cartan anti-involution defined on the
generators by the relations
\begin{equation}
\omega(h_i) = h_i, \qquad \omega(e_i) = f_i, \qquad \omega(f_i) = e_i \label{cai1}
\end{equation}
supplied with the rule $\omega(\hbar) = - \hbar$ implying that $\omega(q) = q^{-1}$. Note that we use
such normalization of the root vectors that for any $\gamma \ne m \delta$ we have
\[
[e_\gamma, f_\gamma] = \frac{q^{h_\gamma} - q^{-h_\gamma}}{q - q^{-1}},
\]
where $h_\gamma = \sum_i m_i h_i$ if $\gamma = \sum_i m_i \alpha_i$. It appears that, as in the case
of $U(\gothg'(A))$, the root vectors corresponding to all roots from $\triangle(A)$ together with the
Cartan generators $h_i$ generate a Poincar\'e--Birkhoff--Witt basis of $U_\hbar(\gothg'(A))$.

\section{\texorpdfstring{Universal $R$-matrix}{Universal R-matrix}}
\label{teh}

We consider the case of a quasi-triangular Hopf algebra, see, for example,~\cite{ChaPre95}. The
corresponding universal $R$-matrix $\calR$ satisfies the Yang--Baxter equation
\begin{equation}
\calR_{12} \, \calR_{13} \, \calR_{23} = \calR_{23} \, \calR_{13} \, \calR_{12}. \label{ybe1}
\end{equation}
and serves as basic object for the construction and investigation of integrable models. In
particular, let for any $\zeta \in \bbC^\times$ a representation $\varphi_\zeta$ of
$U_\hbar(\gothg'(A))$ be given. Then the parameter dependent $R$-matrix
\[
R(\zeta_{1} | \zeta_{2}) = \varphi_{\zeta_1} \otimes \varphi_{\zeta_2} (\calR)
\]
satisfies the Yang-Baxter equation of the form
\begin{equation*}
R_{12}(\zeta_1 | \zeta_2) \, R_{13}(\zeta_1 | \zeta_3) \, R_{23}(\zeta_2 | \zeta_3)
= R_{23}(\zeta_2 | \zeta_3) \, R_{13}(\zeta_1 | \zeta_3) \, R_{12}(\zeta_1 | \zeta_2).
\end{equation*}
Usually, one chooses the representation in such a way that $R(\zeta_1 | \zeta_2)$ depends only on
combination $\zeta_1^{\mathstrut} \zeta_2^{-1}$. This allows one to introduce the $R$-matrix $R(z)$
depending on a single parameter, so that
\[
R(\zeta_1 | \zeta_2) = R(\zeta_1^{\mathstrut} \zeta_2^{-1}).
\]
Now the Yang--Baxter equation takes its original form
\begin{equation}
R_{12}(\zeta_{12}) \, R_{13}(\zeta_{13}) \, R_{23}(\zeta_{23})
= R_{23}(\zeta_{23}) \, R_{13}(\zeta_{13}) \, R_{12}(\zeta_{12}). \label{ybe2}
\end{equation}
Here and below we use the notation $\zeta_{ij}^{\mathstrut} = \zeta_i^{\mathstrut} \zeta_j^{-1}$.

An explicit construction of the universal $R$-matrix for quantum groups was proposed by Khoroshkin and
Tolstoy \cite{TolKho92}. For the case under consideration it looks as follows \cite{KhoTol92}.

First of all, we have to choose a normal ordering \cite{LezSav74, AshSmiTol79} of roots from
$\triangle_+(A)$. In fact there are only two possibilities. We use the ordering where the roots
$\alpha + m \delta$ and $2 \alpha + (2 m + 1) \delta$ come first as
\begin{equation*}
\alpha, \hskip2mm 2\alpha + \delta, \hskip2mm \alpha + \delta,
\hskip2mm 2\alpha + 3\delta, \hskip2mm \alpha + 2\delta, \hskip2mm
2\alpha + 5\delta, \hskip2mm \alpha + 3\delta, \hskip2mm 2\alpha + 7\delta,
\hskip2mm \ldots \hskip2mm,
\end{equation*}
then the roots $m \delta$ come as
\begin{equation*}
\delta, \hskip2mm 2\delta, \hskip2mm 3\delta,
\hskip2mm 4\delta, \hskip2mm 5\delta, \hskip2mm 6\delta, \hskip2mm
7\delta, \hskip2mm \ldots \hskip2mm,
\end{equation*}
and finally the roots $\delta - \alpha + m \delta$ and $\delta - 2 \alpha + 2 m \delta$ come in the
order
\begin{equation*}
\ldots \hskip 2mm, \hskip2mm \delta - 2\alpha + 6\delta, \hskip2mm \delta - \alpha + 2\delta,
\hskip2mm \delta - 2\alpha + 4\delta, \hskip2mm \delta - \alpha + \delta, \hskip2mm \delta -
2\alpha + 2\delta, \hskip2mm \delta - \alpha, \hskip2mm \delta - 2\alpha.
\end{equation*}
Gathering, we can write
\[
\gamma + k\delta \prec m\delta \prec (\delta - \gamma) + \ell\delta,
\]
where $\gamma = \alpha, 2\alpha$. Another ordering is the reverse to this one.

Now we have all the ingredients needed to construct the universal $R$-matrix. According to
Khoroshkin and Tolstoy, it has the form
\[
\calR = \calR_{\prec \delta} \, \calR_{\sim \delta} \, \calR_{\succ \delta} \, \calK.
\]
The first factor is the product of the
$q$-exponentials
\begin{equation}
\calR_{\gamma, \, m} = \exp_{q_{\gamma}} \left( (q - q^{-1}) \,
e_{\gamma + m \delta} \otimes f_{\gamma + m \delta} \right), \label{rgm}
\end{equation}
where $\gamma = \alpha, \, 2 \alpha$, $m \in \bbZ_{\ge 0}$. Here we use the notation
\[
q_{\gamma} = q^{-(\gamma,\gamma)}
\]
and understand the $q$-exponential as the series
\[
\exp_q(x) = 1 + x + \frac{x^2}{(2)_q!} + \ldots + \frac{x^n}{(n)_q!} + \ldots
\]
where
\[
(n)_q! = (n)_q \, (n-1)_q \cdots (2)_q \, (1)_q, \qquad (n)_q = \frac{q^n - 1}{q - 1}.
\]
The order of the factors in $\calR_{\prec \delta}$ coincides with the chosen normal order of the
roots $\gamma + m \delta$. The second factor is
\begin{equation}
\calR_{\sim \delta} = \exp \left( (q - q^{-1}) \sum_{m > 0} b^{-1}_{m} \, e_{m \delta}
\otimes f_{m \delta} \right), \label{rpd}
\end{equation}
where
\begin{equation}
b_{m} = \frac{[m]_q}{m} \left( q^m - (-1)^m + q^{-m} \right). \label{bn}
\end{equation}
Note here that
\[
[e_{\alpha + m\delta} , e_{n\delta}]_q = b_n \, e_{\alpha + (m+n)\delta},
\]
as it used to be for the untwisted case \cite{KhoTol92}. The factor $\calR_{\succ \delta}$ is the
product of the $q$-exponentials
\begin{equation}
\calR_{\delta - \gamma, \, m} = \exp_{q_{\gamma}} \left( (q - q^{-1}) \, e_{(\delta - \gamma) + m
\delta} \otimes f_{(\delta - \gamma) + m \delta} \right), \label{rdmgm}
\end{equation}
where $\gamma = \alpha, 2 \alpha$ and $m \in \bbZ_{\ge 0}$. The order of the factors in $\calR_{\succ \delta}$
coincides with the chosen normal
order of the roots $(\delta - \gamma) + m \delta$. Finally, for the factor $\calK$ we have the
expression
\[
\calK = \exp \left( {\hbar \, h_{\alpha} \otimes h_{\alpha}} \right).
\]

\section{Finite-dimensional representation}

Given $\zeta \in \bbC^\times$, define the three-dimensional representation $\varphi_\zeta$ of the
quantum group $U_\hbar(\gothg'(A)$ by the relations
\begin{align}
& \varphi_\zeta (e_{\alpha}) = \zeta^{s_1} \left( E_{12} + E_{23} \right), \qquad  \varphi_\zeta
(e_{\delta - 2\alpha}) = \zeta^{s_0} \, ([2]_q)^{1/2} \, E_{31}, \label{varz1} \\
& \varphi_\zeta (h_{\alpha}) = E_{11} - E_{33}, \qquad \qquad  \varphi_\zeta
(h_{\delta - 2\alpha}) = -2 E_{11} + 2 E_{33}, \label{varz2}\\
& \varphi_\zeta (f_{\alpha}) = \zeta^{-s_1} \left( E_{21} + E_{32} \right), \qquad  \varphi_\zeta
(f_{\delta - 2\alpha}) = \zeta^{-s_0} \, ([2]_q)^{1/2} \, E_{13}, \label{varz3}
\end{align}
where $s_i$, $i = 0, 1$ are some integers, $i = 0, 1$, and the $3 \times 3$ matrix units $E_{ij}$
are defined as
\[
(E_{ij})_{mn} = \delta_{im} \, \delta_{jn}.
\]
The basic property following from this definition and used in what follows is given by the relation
\[
E_{ij} \, E_{kl} = \delta_{jk} \, E_{il}.
\]
Note that $\varphi_\zeta(h_{\delta-2\alpha}) + 2\varphi_\zeta(h_{\alpha}) = 0$. If we have an
expression for $\varphi_\zeta(a)$, where $a \in U_\hbar(\gothg'(A))$, in order to obtain the
expression for $\varphi_\zeta(\omega(a))$, we should simply take the transpose of
$\varphi_\zeta(a)$, also changing the deformation and spectral parameters as $q$ to $q^{-1}$ and
$\zeta$ to $\zeta^{-1}$. We denote this operation by $\Omega_3$, where the index corresponds to the
rank of the matrix it acts on. We will use similar operations for matrices of different ranks.

Now, using the recursive relations for the higher root vectors as given above and equations
(\ref{varz1})--(\ref{varz3}), we obtain\footnote{Here and in what follows, we use instead of the
integers $s_0$ and $s_1$ the integers $s = s_0 + 2s_1$ and $s_1$.}
\begin{align}
& \varphi_{\zeta} (e_{\alpha + m\delta}) = q^{-m} \, \zeta^{s_1 + ms}
\left( (-1)^m E_{12} + q^{-m} E_{23} \right), \label{hivec1a}\\
& \varphi_{\zeta} (f_{\alpha + m\delta}) = q^{m} \, \zeta^{-s_1 - ms} \left( (-1)^m E_{21} + q^m
E_{32} \right), \label{hivec2a}
\end{align}
\begin{align}
& \varphi_{\zeta} (e_{\delta - \alpha + m\delta}) = -q^{-m} \, \zeta^{s-s_1 + ms} \left( (-1)^{m+1}
E_{21} + q^{-m-2}
E_{32} \right), \label{hivec3a}\\
& \varphi_{\zeta} (f_{\delta - \alpha + m\delta}) = -q^{m} \, \zeta^{-(s-s_1) - ms} \left(
(-1)^{m+1} E_{12} + q^{m+2} E_{23} \right), \label{hivec4a}
\end{align}
\begin{align}
& \varphi_{\zeta} (e_{2\alpha + (2m+1)\delta}) = (-1)^m \, q^{-3m-1} \, \zeta^{2s_1 + (2m+1)s} \,
 ([2]_q)^{1/2} \, E_{13},
\label{hivec5a} \\
& \varphi_{\zeta} (f_{2\alpha + (2m+1)\delta}) = (-1)^m \, q^{3m+1} \, \zeta^{-2s_1 - (2m+1)s} \,
 ([2]_q)^{1/2} \, E_{31}, \label{hivec6a}
\end{align}
\begin{align}
& \varphi_{\zeta} (e_{\delta - 2\alpha + 2m\delta}) = (-1)^m \, q^{-3m} \, \zeta^{s-2s_1 + 2ms}
\, ([2]_q)^{1/2} \, E_{31}, \label{hivec7a} \\
& \varphi_{\zeta} (f_{\delta - 2\alpha + 2m\delta}) = (-1)^m \, q^{3m} \, \zeta^{-(s-2s_1) - 2ms}
\, ([2]_q)^{1/2} \, E_{13}, \label{hivec8a}
\end{align}
where $m = 0, 1, 2, \ldots$. For the primed positive imaginary root vectors we have the following
expressions:
\begin{align}
\varphi_{\zeta} (e^{\prime}_{m\delta}) = q^{-m+1} \, \zeta^{ms} \left( (-1)^{m-1} \, E_{11} +
   [(-1)^m q^{-1} - q^{-m-1}] \, E_{22} + q^{-m-2} \, E_{33} \right), & \label{hivec9a} \\
\varphi_{\zeta} (f^{\prime}_{m\delta}) = q^{m-1} \, \zeta^{-ms} \left( (-1)^{m-1} \, E_{11} +
  [(-1)^m q - q^{m+1}] \, E_{22} + q^{m+2} \, E_{33} \right). & \label{hivec10a}
\end{align}
This allows us to pass to unprimed quantities via the relation
\[
(q-q^{-1}) \varphi_\zeta (e_\delta(x)) = \varphi_\zeta (\log[1 + (q -
q^{-1})e^{\prime}_\delta(x)]),
\]
where
\[
e^{\prime}_\delta(x) = \sum_{m>0} e^{\prime}_{m\delta} x^{-m}, \qquad e_\delta(x) = \sum_{m>0}
e_{m\delta} x^{-m}.
\]
The corresponding expressions for the unprimed generators $f_{m\delta}$ can be
found then by applying the Cartan anti-involution. After some calculations we find
\begin{align}
\varphi_{\zeta} (e_{m\delta}) = \frac{[m]_q}{m} \zeta^{ms} \left( (-1)^{m-1} \, E_{11} +
  [(-1)^m q^{-2m} - q^{-m}] \, E_{22} + q^{-3m} \, E_{33} \right), & \label{emd} \\
\varphi_{\zeta} (f_{m\delta}) = \frac{[m]_q}{m} \zeta^{-ms} \left( (-1)^{m-1} \, E_{11} +
  [(-1)^m q^{2m} - q^{m}] \, E_{22} + q^{3m} \, E_{33} \right), & \label{fmd}
\end{align}
where $m$ runs over the set of all positive integers.

\section{\texorpdfstring{$R$-matrix}{R-matrix}}
\label{urm}

In this section, we construct the $R$-matrix corresponding to the representation $\varphi_\zeta$ defined
in the previous section. As usual, the most
cumbersome part of the calculations is about the factor $\varphi_{\zeta_1} \otimes
\varphi_{\zeta_2} (\calR_{\sim \delta})$. We obtain the following diagonal matrix:
\begin{multline*}
\varphi_{\zeta_1} \otimes \varphi_{\zeta_2} (\calR_{\sim \delta}) = \rme^{\lambda(q \zeta_{12}^s)
- \lambda(q^{-1} \zeta_{12}^s)}
\left[ E_{11} \otimes E_{11} + \frac{1 + q\zeta^s_{12}}{1 + q^{-1}\zeta^s_{12}} \, E_{22} \otimes
E_{22} + E_{33} \otimes E_{33} \right. \\ \left. + \frac{1 - q^2\zeta^s_{12}}{1 - \zeta^s_{12}}
\left( E_{11} \otimes E_{22} + E_{22} \otimes E_{33} \right) + \frac{1 - \zeta^s_{12}}{1 -
q^{-2}\zeta^s_{12}} \left( E_{22} \otimes E_{11} + E_{33} \otimes E_{22} \right) \right.  \\
\left. + \frac{1 - q^2\zeta^s_{12}}{1 - \zeta^s_{12}} \, \frac{1 + q^3\zeta^s_{12}}{1 +
q\zeta^s_{12}} \, E_{11} \otimes E_{33} + \frac{1 - \zeta^s_{12}}{1 - q^{-2}\zeta^s_{12}} \,
\frac{1 + q^{-1}\zeta^s_{12}}{1 + q^{-3}\zeta^s_{12}} \, E_{33} \otimes E_{11} \right],
\end{multline*}
where
\[
\lambda(\zeta) = \sum_{m > 0} \frac{1}{q^{m} - (-1)^{m} + q^{-m}}
\frac{\zeta^m}{m}.
\]
A useful property of the transcendental function $\lambda$ is that
\begin{equation}
\lambda(q\zeta) - \lambda(-\zeta) + \lambda(q^{-1}\zeta) = - \log (1 - \zeta). \label{lbr}
\end{equation}

Further, using equations (\ref{hivec1a}), (\ref{hivec2a}) and (\ref{hivec5a}), (\ref{hivec6a}) we
see that $\varphi_{\zeta_1} \otimes \varphi_{\zeta_2} (\calR_{\alpha, \, m})$ commute with
$\varphi_{\zeta_1} \otimes \varphi_{\zeta_2} (\calR_{2\alpha, \, n})$ for any $m$ and $n$, where
$\calR_{\alpha, \, m}$ and $\calR_{2\alpha, \, n}$ are factors in $\calR_{\prec \delta}$
corresponding to the roots $\alpha + m\delta$ and $2\alpha + (2n+1)\delta$, respectively.
Similarly, using equations (\ref{hivec3a}), (\ref{hivec4a}) and (\ref{hivec7a}), (\ref{hivec8a}) we
see that $\varphi_{\zeta_1} \otimes \varphi_{\zeta_2} (\calR_{\delta - 2\alpha, \, m})$ commute
with $\varphi_{\zeta_1} \otimes \varphi_{\zeta_2} (\calR_{\delta - \alpha, \, n})$ for any $m$ and
$n$. Here $\calR_{\delta - 2\alpha, \, m}$ and $\calR_{\delta - \alpha, \, n}$ are factors present
in $\calR_{\succ \delta}$ corresponding to the roots $\delta - 2\alpha + 2m\delta$ and $\delta -
\alpha + n\delta$, respectively. Therefore, we can rearrange the factors entering
$\varphi_{\zeta_1} \otimes \varphi_{\zeta_2} (\calR_{\prec \delta})$ in the definition of the
universal $R$-matrix, so that the factors corresponding to the roots $\alpha + m\delta$ will come
first, and then come the factors corresponding to the roots $2\alpha + (2m+1)\delta$. Similarly, we
can rearrange the factors entering $\varphi_{\zeta_1} \otimes \varphi_{\zeta_2} (\calR_{\succ
\delta})$ in the definition of the universal $R$-matrix in such a way that the factors
corresponding to the roots $\delta - 2\alpha + 2m\delta$ come first, and only then come the factors
corresponding to the roots $\delta - \alpha + m\delta$. The same useful rearrangement turns out to
be also valid for the matter of constructing the $L$-operators.

Hence, in the case under consideration we can write
\[
\varphi_{\zeta_1} \otimes \varphi_{\zeta_2} (\calR_{\prec \delta}) =  \varphi_{\zeta_1} \otimes
\varphi_{\zeta_2} (\calR_{\alpha} \calR_{2\alpha}), \qquad \varphi_{\zeta_1} \otimes
\varphi_{\zeta_2} (\calR_{\succ \delta}) = \varphi_{\zeta_1} \otimes \varphi_{\zeta_2}
(\calR_{\delta - 2\alpha} \calR_{\delta - \alpha}),
\]
where the factors entering this definition are given by the expressions
\begin{align*}
& \varphi_{\zeta_1} \otimes \varphi_{\zeta_2} (\calR_{\alpha}) = \prod^{\curvearrowright}_{m \ge 0}
\exp_{q_{\alpha}} \left[ (q-q^{-1}) \, \varphi_{\zeta_1}(e_{\alpha +
m\delta}) \otimes \varphi_{\zeta_2}(f_{\alpha + m\delta})
\right], \\
& \varphi_{\zeta_1} \otimes \varphi_{\zeta_2} (\calR_{2\alpha}) = \prod^{\curvearrowright}_{m \ge
0} \exp_{q_{2\alpha}} \left[ (q-q^{-1}) \,
\varphi_{\zeta_1}(e_{2\alpha + (2m+1)\delta}) \otimes \varphi_{\zeta_2}(f_{2\alpha + (2m+1)\delta})
\right],
\end{align*}
and
\begin{align*}
& \varphi_{\zeta_1} \otimes \varphi_{\zeta_2}(\calR_{\delta - 2\alpha}) =
\prod^{\curvearrowleft}_{m \ge 0} \exp_{q_{2\alpha}} \left[ (q-q^{-1}) \,
\varphi_{\zeta_1}(e_{\delta - 2\alpha + 2m\delta}) \otimes
\varphi_{\zeta_2}(f_{\delta - 2\alpha + 2m\delta}) \right], \\
& \varphi_{\zeta_1} \otimes \varphi_{\zeta_2} (\calR_{\delta - \alpha}) =
\prod^{\curvearrowleft}_{m \ge 0} \exp_{q_{\alpha}} \left[ (q-q^{-1}) \,
\varphi_{\zeta_1}(e_{\delta - \alpha + m\delta}) \otimes
\varphi_{\zeta_2}(f_{\delta - \alpha + m\delta}) \right].
\end{align*}

Using the expressions for $\varphi_{\zeta_1} \otimes \varphi_{\zeta_2} (\calR_{\alpha})$,
$\varphi_{\zeta_1} \otimes \varphi_{\zeta_2} (\calR_{2\alpha})$ and $\varphi_{\zeta_1} \otimes
\varphi_{\zeta_2} (\calR_{\delta-2\alpha})$, $\varphi_{\zeta_1} \otimes \varphi_{\zeta_2}
(\calR_{\delta-\alpha})$ presented in Appendix \ref{clcRmat}, we obtain by the respective matrix multiplication
\begin{multline*}
\varphi_{\zeta_1} \otimes \varphi_{\zeta_2} (\calR_{\prec \delta}) = I +
\frac{q-q^{-1}}{1-\zeta^s_{12}} \, \zeta^{s_1}_{12} \, \left(
E_{12} \otimes E_{21} + E_{23} \otimes E_{32} \right) \\
+ \frac{q-q^{-1}}{1+q\zeta^s_{12}} \, \zeta^{s_1}_{12} \, E_{12} \otimes E_{32} +
\frac{q-q^{-1}}{1+q^{-1}\zeta^s_{12}} \, \zeta^{s_1}_{12} \, E_{23} \otimes E_{21} \\ +
\frac{(q-q^{-1})(q-1+(q+q^{-1})q^{-1}\zeta^s_{12})}{(1-\zeta^s_{12})(1+q^{-1}\zeta^s_{12})} \,
\zeta^{2s_1}_{12} \, E_{13} \otimes E_{31}
\end{multline*}
and
\begin{multline*}
\varphi_{\zeta_1} \otimes \varphi_{\zeta_2} (\calR_{\succ \delta}) = I +
\frac{(q-q^{-1})\zeta^{s-s_1}_{12}}{1-\zeta^s_{12}} \left(
E_{21} \otimes E_{12} + E_{32} \otimes E_{23} \right) \\
- \frac{(q-q^{-1})q^2\zeta^{s-s_1}_{12}}{1+q\zeta^s_{12}} \, E_{21} \otimes E_{23} -
\frac{(q-q^{-1})q^{-2}\zeta^{s-s_1}_{12}}{1+q^{-1}\zeta^s_{12}} \, E_{32} \otimes E_{12} \\
+ \frac{(q-q^{-1})(q+q^{-1}+(q^{-1}-1)q^{-1}\zeta^s_{12})\zeta^{s-2s_1}_{12}}
{(1-\zeta^s_{12})(1+q^{-1}\zeta^s_{12})} \, E_{31} \otimes E_{13}.
\end{multline*}
For the simplest part of the calculations, the last factor in the
definition of the universal $R$-matrix, we find the following
diagonal matrix
\begin{multline*}
\varphi_{\zeta_1} \otimes \varphi_{\zeta_2} (\calK) =  q \, E_{11} \otimes E_{11} + E_{11} \otimes
E_{22} + q^{-1} \, E_{11} \otimes E_{33}
\\ + E_{22} \otimes E_{11} + E_{22} \otimes E_{22} + E_{22} \otimes E_{33}  \\
+ q^{-1} \, E_{33} \otimes E_{11} + E_{33} \otimes E_{22} + q \,
E_{33} \otimes E_{33}.
\end{multline*}

Now, multiplying all the factors in the given order, we finally obtain
\[
\varphi_{\zeta_1} \otimes \varphi_{\zeta_2} (\calR)
= \rme^{\lambda(q \zeta_{12}^s) - \lambda(q^{-1} \zeta_{12}^s)} \,
R(\zeta_{12}),
\]
where
\begin{multline*}
R(\zeta) =  q \, E_{11} \otimes E_{11} + \rho(\zeta) \left[ E_{11}
\otimes E_{22} + a(\zeta)
\, E_{12} \otimes E_{21} \right. \\*[.5em] \left. + \Omega_9 \left(a(\zeta)
\, E_{12} \otimes E_{21}\right) + E_{22} \otimes E_{11} \right] \\*[.5em]
\sigma(\zeta) \left[ E_{11} \otimes E_{33} + q \, b(\zeta) \, E_{12}
\otimes E_{32} + c(\zeta) \, E_{13} \otimes E_{31} + b(\zeta) \,
E_{23} \otimes E_{21} \right. \\*[.5em]
+ \Omega_9 (q \, b(\zeta) \, E_{12} \otimes E_{32} + c(\zeta) \, E_{13}
\otimes E_{31} + b(\zeta) \, E_{23} \otimes E_{21}) \\[.5em] \left. +
d(\zeta) \, E_{22} \otimes E_{22} + E_{33} \otimes E_{11} \right]
\\*[.5em]
+ \rho(\zeta) \left[ E_{22} \otimes E_{33} + a(\zeta) \, E_{23} \otimes E_{32} \right. \\* \left. +
\Omega_9 (a(\zeta) \, E_{23} \otimes E_{32}) + E_{33} \otimes E_{22} \right] + q \, E_{33} \otimes
E_{33}.
\end{multline*}
Here we use the notations
\begin{gather*}
a(\zeta) = \frac{(q - q^{-1})}{1 - \zeta^s} \, \zeta^{s_1}, \qquad b(\zeta) = \frac{(q - q^{-1})}{1
+ q^{-1}\zeta^s} \, \zeta^{s_1}, \\ c(\zeta) =
\frac{(q-q^{-1})(q-1+(q+q^{-1})q^{-1}\zeta^s)}{(1-\zeta^s)(1+q^{-1}\zeta^s)} \, \zeta^{2s_1}, \\[.5em]
d(\zeta) = \frac{q+(q-1)(q-q^{-1}+q^{-3})\zeta^s - q^{-2}\zeta^{2s}}{(1-\zeta^s)(1+q^{-1}\zeta^s)},
\\[.5em]
\rho(\zeta) = \frac{1-\zeta^s}{1-q^{-2}\zeta^s}, \qquad \sigma(\zeta) =
\frac{q^{-1}(1-\zeta^s)(1+q^{-1}\zeta^s)}{(1-q^{-2}\zeta^s)(1+q^{-3}\zeta^s)}.
\end{gather*}
Choosing $s_0 = 1$, $s_1 = 0$ we recover the $R$-matrix presented in \cite{KhoTol92}. To compare
with the original result of \cite{IzeKor81}, one should apply certain similarity transformation and
also adjust the respective parametrization, see, for example, \cite{KulSkl82, MezNep92}. Here we
have that
\[
\Omega_1 (d(\zeta)) = d(\zeta), \qquad \Omega_1 (\rho(\zeta)) = q^{-2} \rho(\zeta), \qquad
\Omega_1 (\sigma(\zeta)) = q^{-2} \sigma(\zeta),
\]
and we take into account that we have
\[
\Omega_9 (E_{ij} \otimes E_{mn}) = E_{ji} \otimes E_{nm}.
\]
The explicit matrix form of the $R$-matrix for the considered representation is given in Figure
\ref{fig1}.
\begin{figure}
\[
\psset{xunit=3.0em, yunit=2.4em} R = \left( \raise 0.5\psyunit
\hbox{\begin{pspicture}(9.2,4.4)(.8,.4) \rput(1,4){$q$} \rput(2,3){$\rho$} \rput(4,3){$\rho a$}
\rput(2,1){$\rho \Omega_1(a)$} \rput(4,1){$\rho$} \rput(3,2){$\sigma$} \rput(5,2){$q \sigma b$}
\rput(7,2){$\sigma c$} \rput(3,0){$q^{-1} \sigma \Omega_1(b)$} \rput(5,0){$\sigma d$}
\rput(7,0){$\sigma b$} \rput(3,-2){$\sigma \Omega_1(c)$} \rput(5,-2){$\sigma \Omega_1(b)$}
\rput(7,-2){$\sigma$} \rput(6,-1){$\rho$} \rput(8,-1){$\rho a$} \rput(6,-3){$\rho \Omega_1(a)$}
\rput(8,-3){$\rho$} \rput(9,-4){$q$} \psline(2.15,3)(3.75,3) \psline(2,2.8)(2,1.2)
\psline(4,2.8)(4,1.25) \psline(2.5,1)(3.8,1) \psline(6.15,-1)(7.75,-1) \psline(6,-1.2)(6,-2.8)
\psline(8,-1.2)(8,-2.75) \psline(6.5,-3)(7.8,-3) \psline(3.15,2)(4.65,2) \psline(5.3,2)(6.75,2)
\psline(3,1.8)(3,0.3) \psline(5,1.8)(5,0.3) \psline(7,1.8)(7,0.3) \psline(3,-0.2)(3,-1.8)
\psline(5,-.2)(5,-1.8) \psline(7,-0.2)(7,-1.8) \psline(3.75,0)(4.75,0) \psline(5.25,0)(6.75,0)
\psline(3.5,-2)(4.48,-2) \psline(5.5,-2)(6.8,-2)
\end{pspicture}} \right)
\]
\caption{} \label{fig1}
\end{figure}
It is clear, in particular, that diagonalizing this matrix means an independent diagonalization of
two its $2 \times 2$ sub-matrices with $\rho$ and one $3 \times 3$ block with $\sigma$. Here we
obtain three different eigenvalues $\{q, \; q\frac{\displaystyle q^2\zeta^s - 1}{\displaystyle q^2
- {\zeta^s}}, \; q \frac{\displaystyle q^3{\zeta^s} + 1}{\displaystyle q^{3}+{\zeta^s}}\}$, having
multiplicities $\{5, \, 3, \, 1\}$, respectively.\footnote{Compare with the untwisted case
associated with $A^{(1)}_2$, where one has only two different eigenvalues $\{ 1, \;
\frac{q^2{\zeta^s} - 1}{q^2-{\zeta^s}}\}$ with multiplicities $\{6, \, 3\}$, respectively.}

The $R$-matrices corresponding to different values of $s = s_0 + 2s_1$ and $s_1$ are related by a
change of the spectral parameter and a gauge transformation \cite{BraGouZhaDel94,
BooGoeKluNirRaz10}. In the case under consideration we have the following relation:\footnote{It is
implied here that $R(\zeta) = R^{(s,s_1)}(\zeta)$, $R^{(1,0)}(\zeta^s) = R^{(s,0)}(\zeta)$, and a
similar convention will be used for the $L$-operators.}
\[
R^{(s,s_1)}(\zeta_{12}) = [G(\zeta_1) \otimes G(\zeta_2)] \, R^{(1,0)}(\zeta^s_{12}) \, [G(\zeta_1)
\otimes G(\zeta_2)]^{-1},
\]
where
\begin{equation}
G(\zeta) = \left( \begin{array}{ccc} \zeta^{s_1} & 0 & 0 \\
                                               0 & 1 & 0 \\
                                               0 & 0 & \zeta^{-s_1} \end{array} \right)
\label{Gz}
\end{equation}
and $\zeta_{12}$ denotes the ratio $\zeta_1/\zeta_2$.

Note finally that the expression for the $R$-matrix remarkably factorizes. Indeed, the expression
for the universal $R$-matrix can be written as
\[
\calR = ( \calR_{\prec \delta} ) (\calR_{\sim \delta} \, \calK ) ( \calK^{-1} \calR_{\succ \delta}
\calK ).
\]
Observing that the relation
\[
\varphi_{\zeta_1} \otimes \varphi_{\zeta_2} (\calK^{-1} \calR_{\succ \delta} \calK) = \Omega_9 \circ
\varphi_{\zeta_1} \otimes \varphi_{\zeta_2} (\calR_{\prec \delta})
\]
holds in the case under consideration, we can thus write
\[
R = R_+ \, R_0 \, R_-,
\]
where we have denoted the upper-triangular, diagonal and
lower-triangular factors as
\begin{multline*}
R_+(\zeta_{12}) = \varphi_{\zeta_1} \otimes \varphi_{\zeta_2} (\calR_{\alpha} \calR_{2\alpha}) = I
+ \frac{q-q^{-1}}{1-\zeta^s_{12}} \, \zeta^{s_1}_{12} \left(
E_{12} \otimes E_{21} + E_{23} \otimes E_{32} \right) \\
+ \frac{q-q^{-1}}{1+q\zeta^s_{12}} \, \zeta^{s_1}_{12} \, E_{12} \otimes E_{32} +
\frac{q-q^{-1}}{1+q^{-1}\zeta^s_{12}} \, \zeta^{s_1}_{12} \, E_{23} \otimes E_{21} \\ +
\frac{(q-q^{-1})(q-1+(q+q^{-1})q^{-1}\zeta^s_{12})}
{(1-\zeta^s_{12})(1+q^{-1}\zeta^s_{12})} \,
\zeta^{2s_1}_{12} \, E_{13} \otimes E_{31},
\end{multline*}
\begin{multline*}
R_0(\zeta_{12}) = \rme^{- \lambda(q \zeta_{12}^s) + \lambda(q^{-1} \zeta_{12}^s)} \varphi_{\zeta_1}
\otimes \varphi_{\zeta_2}
(\calR_{\sim \delta} \calK) = q \, E_{11} \otimes E_{11} \\
 + \frac{1 - q^2\zeta^s_{12}}{1 - \zeta^s_{12}} \, E_{11} \otimes
E_{22} + \frac{(1 - q^2\zeta^s_{12})(1 + q^3\zeta^s_{12})}{(1 - \zeta^s_{12})(1 +
q\zeta^s_{12})} \, q^{-1} \, E_{11} \otimes E_{33}  \\
+ \frac{1 - \zeta^s_{12}}{1 - q^{-2}\zeta^s_{12}} \, E_{22} \otimes E_{11} + \frac{1 +
q\zeta^s_{12}}{1 + q^{-1}\zeta^s_{12}} \, E_{22} \otimes E_{22} + \frac{1 - q^2\zeta^s_{12}}{1 -
\zeta^s_{12}} \, E_{22} \otimes E_{33}  \\
+ \frac{(1 - \zeta^s_{12})(1 + q^{-1}\zeta^s_{12})}{(1 - q^{-2}\zeta^s_{12})(1 +
q^{-3}\zeta^s_{12})} \, q^{-1} \, E_{33} \otimes E_{11} + \frac{1 - \zeta^s_{12}}{1 -
q^{-2}\zeta^s_{12}} \, E_{33} \otimes E_{22} + q \, E_{33} \otimes E_{33},
\end{multline*}
\begin{multline*}
R_-(\zeta_{12}) = \varphi_{\zeta_1} \otimes \varphi_{\zeta_2} (\calK^{-1} \calR_{\delta-2\alpha}
\calR_{\delta-\alpha} \calK) = I + \frac{(q-q^{-1})\zeta^{s-s_1}_{12}}{1-\zeta^s_{12}} \left(
E_{21} \otimes E_{12} + E_{32} \otimes E_{23} \right) \\
- \frac{(q-q^{-1})q\zeta^{s-s_1}_{12}}{1+q\zeta^s_{12}} \, E_{21} \otimes E_{23} -
\frac{(q-q^{-1})q^{-1}\zeta^{s-s_1}_{12}}{1+q^{-1}\zeta^s_{12}} \, E_{32} \otimes E_{12} \\
+ \frac{(q-q^{-1})(q+q^{-1}+(q^{-1}-1)q^{-1}\zeta^s_{12})\zeta^{s-2s_1}_{12}}
{(1-\zeta^s_{12})(1+q^{-1}\zeta^s_{12})} \, E_{31} \otimes E_{13},
\end{multline*}
respectively. Here we also have the Cartan anti-involution relations
\[
\Omega_9 (R_+(\zeta)) = R_-(\zeta), \qquad \Omega_9 (R_0(\zeta)) = q^{-2} \, R_0(\zeta),
\]
implying that
\[
\Omega_9 (R(\zeta)) = q^{-2} R(\zeta).
\]

\section{The spin-chain Hamiltonian}
\label{sch}

The transfer-matrix of the system based on $N$ sites is given by the relation\footnote{Also a
certain twist operator can be introduced in the definition of the transfer-matrix
corresponding to specific boundary conditions.}
\[
T(\zeta | \xi_1,\ldots,\xi_N) = \mathrm{tr}_0\left( R_{01}(\zeta/\xi_1) R_{02}(\zeta/\xi_2)
\cdots R_{0N}(\zeta/\xi_N)\right),
\]
where the string of $R_{0k}(\zeta/\xi_k)$ acts in $V_0 \otimes V_1 \otimes \ldots \otimes V_N$ and
the trace is taken over $V_0$ in the given representation. The corresponding Hamiltonian is defined
by the formula
\[
\wt{H} = \left. T^{-1}(\zeta) \, \frac{\rmd T(\zeta)}{\rmd \zeta} \right|_{\zeta = 1},
\]
where $T(\zeta) = T(\zeta | 1, \ldots, 1)$, and allows one to obtain the ground-state energy of the
system, see, for example, \cite{VicRes83,WarBatNie92}.

Direct calculations show that the Hamiltonian in the case under consideration is a sum of four
terms,
\[
\wt{H} = H_{(12)} + H_{(23)} + H_{(31)} + H_{(123)},
\]
where
\begin{gather*}
H_{(12)} = - \frac{1}{q-q^{-1}} \sum^N_{l=1}\left[ E^{(l)}_{12} \, E^{(l+1)}_{21} + E^{(l)}_{21} \,
E^{(l+1)}_{12} - q^{-1} \, E^{(l)}_{11} \, E^{(l+1)}_{22} - q \, E^{(l)}_{22} \, E^{(l+1)}_{11}
\right], \\
H_{(23)} = - \frac{1}{q-q^{-1}} \sum^N_{l=1}\left[ E^{(l)}_{23} \, E^{(l+1)}_{32} + E^{(l)}_{32} \,
E^{(l+1)}_{23} - q^{-1} \, E^{(l)}_{22} \, E^{(l+1)}_{33} - q \, E^{(l)}_{33} \, E^{(l+1)}_{22}
\right],
\end{gather*}
\begin{multline*}
H_{(31)} = - \frac{1}{q-q^{-1}} \frac{[3]_{q^{1/2}}}{[3]_q} \sum^N_{l=1}\left[ E^{(l)}_{31} \,
E^{(l+1)}_{13} + E^{(l)}_{13} \, E^{(l+1)}_{31} \right. \\* \left. - (q^{-1} + q(q-1)) \,
E^{(l)}_{33} \, E^{(l+1)}_{11} - (q - q^{-2}(q-1)) \, E^{(l)}_{11} \, E^{(l+1)}_{33} \right],
\end{multline*}
\begin{multline*}
H_{(123)} = - \frac{[3/2]_{q}}{[3]_q} \sum^N_{l=1}\left[ q^{1/2} \left( q \, E^{(l)}_{12} \,
E^{(l+1)}_{32} - q^{-1} \, E^{(l)}_{32} \, E^{(l+1)}_{12} \right) \right. \\ \left. + q^{-1/2}
\left(q \, E^{(l)}_{21} \, E^{(l+1)}_{23} - q^{-1} \, E^{(l)}_{23} \, E^{(l+1)}_{21} \right) -
(q^{1/2} - q^{-1/2}) [2]_q \, E^{(l)}_{22} \, E^{(l+1)}_{22} \right]
\end{multline*}
in the standard notation, with $s_i$ fixed as $s_0=1$, $s_1=0$.

Recall that for the untwisted case, i.~e. the system associated with $A^{(1)}_{n-1}$, we have the
Hamiltonian also expressed in terms of the respective matrix units,
\[
H_n = - \frac{1}{q-q^{-1}} \sum^N_{l=1}\biggl[ \sum^n_{\substack{i, j = 1 \\ i \neq j}} E^{(l)}_{ij} \,
E^{(l+1)}_{ji} - q^{-1} \sum^n_{i<j} E^{(l)}_{ii} \, E^{(l+1)}_{jj} - q \sum^n_{i>j} E^{(l)}_{ii}
\, E^{(l+1)}_{jj} \biggr].
\]
In the best studied cases, $n=2$ and $n=3$, the Hamiltonian can be represented by means of the
generators of the $1$st fundamental representation of the corresponding finite-dimensional
algebras, $A_1$ and $A_2$, respectively,
\begin{multline*}
H_2 = - \frac{1}{q-q^{-1}} \sum^N_{l=1} \biggl[ E^{(l)} \, F^{(l+1)} + F^{(l)} \, E^{(l+1)} -
\frac{q+q^{-1}}{4} H^{(l)} \, H^{(l+1)} \biggr] \\ + \frac{1}{4} \left( H^{(1)} - H^{(N+1)} \right)
- \frac{N}{4} \frac{q+q^{-1}}{q-q^{-1}},
\end{multline*}
\begin{multline*}
H_3 = - \frac{1}{q-q^{-1}} \sum^N_{l=1} \biggl[ E^{(l)}_\alpha \, F^{(l+1)}_\alpha +
E^{(l)}_{\alpha+\beta} \, F^{(l+1)}_{\alpha+\beta} + E^{(l)}_\beta \, F^{(l+1)}_\beta \\ +
F^{(l)}_\alpha \, E^{(l+1)}_\alpha + F^{(l)}_{\alpha+\beta} \, E^{(l+1)}_{\alpha+\beta} +
F^{(l)}_\beta \, E^{(l+1)}_\beta \\ + \frac{q+q^{-1}}{3}\left( H^{(l)}_\alpha \, H^{(l+1)}_\alpha -
H^{(l)}_\beta \, H^{(l+1)}_\beta \right) + \frac{q}{3} H^{(l)}_\alpha \, H^{(l+1)}_\beta
+ \frac{q^{-1}}{3} H^{(l)}_\beta \, H^{(l+1)}_\alpha \biggr] \\
- \frac{1}{3} \left( H^{(1)}_\alpha - H^{(N+1)}_\alpha + H^{(1)}_\beta - H^{(N+1)}_\beta \right) +
\frac{N}{3} \frac{q+q^{-1}}{q-q^{-1}}.
\end{multline*}
Here we kept the boundary terms and the constant terms.

Now, comparing these expressions with what we have obtained for the twisted $A_2^{(2)}$ case, we
see that the Hamiltonian of the system under consideration is the sum of two pure $A_1^{(1)}$-type
systems' Hamiltonians $H_{(12)}$ and $H_{(23)}$, related to the indices $(12)$ and $(23)$, one more
somewhat shifted $A_1^{(1)}$-type system's Hamiltonian $H_{(31)}$, and an essentially $A_2^{(2)}$
addition presented explicitly by the part $H_{(123)}$. Here, we cannot express the Hamiltonian
$\wt{H}$ in terms of matrices representing the algebra generators, which is in contrast with the
untwisted cases.

\section{\texorpdfstring{$L$-operators in $q$-oscillator representation}{L-operators
in q-oscillator representation}}
\label{lor}

A useful object to investigate the properties of an integrable system is the corresponding
$Q$-operator. According to the modern approach, it is constructed as the trace of some monodromy
type operator constructed, in turn, from an $L$-operator. Here the $L$-operator is obtained by
taking one of the factors of the universal $R$-matrix in an infinite-dimensional representation.
Usually, it is some $q$-oscillator representation \cite{BazLukZam97, BazHibKho02,
BooGoeKluNirRaz10}.

In the case under consideration, $\calR$ is an element of the tensor product of the Borel
subalgebras of the quantum group $U_\hbar(\gothg'(A))$, and thus we have
\[
\calR \in  U_\hbar(\gothb'_+(A)) \otimes U_\hbar(\gothb'_-(A)) \subset
U_\hbar(\gothg'(A)) \otimes U_\hbar(\gothg'(A)).
\]
The two Borel subalgebras here are unital associative algebras generated by the elements $h_i$,
$e_i$ and $h_i$, $f_i$ respectively. To construct an $L$-operator it therefore suffices to have
representations of the Borel subalgebras.

\subsection{Resolving the Serre relations}
\label{rsr}

In this section we construct $L$-operators based on the $q$-deformed oscillator algebra defined as
an associative algebra $\rmOsc_\hbar$ with generators $a$, $a^\dagger$ and $D$ subject to the
relations\footnote{Instead of the naturally $q$-deformed oscillators, defined by the relations
$[{\calN},b] = -b$, $[{\calN},b^\dagger] = b^\dagger$, $b b^\dagger = [{\calN}+1]_q$, $b^\dagger b
= [{\calN}]_q$, we use slightly different objects used earlier in
\cite{BooJimMiwSmi09,BooJimMiwSmiTak07,BooJimMiwSmiTak09}. The latter are related with the former
as $D = {\calN}$, $a = - (q - q^{-1}) b q^{\calN}$, $a^\dagger = b^\dagger$.}
\begin{align*}
& [D,a] = -a, \qquad \qquad [D,a^\dagger] = a^\dagger, \\
& a a^\dagger = 1 - q^2 q^{2D}, \qquad a^\dagger a = 1 - q^{2D}.
\end{align*}
The transformations
\begin{equation}
a \to \kappa \, a \, q^{\xi D}, \qquad a^\dagger \to \kappa^{-1} q^{-\xi D} a^\dagger, \qquad D \to
D \label{2pag}
\end{equation}
form a two-parametric group of automorphisms of $\rmOsc_\hbar$. One can use these transformations
to obtain different $L$-operators, however, the trace used in the definition of $Q$-operators is
invariant with respect to the action of this automorphism group
\cite{BazLukZam97,BooJimMiwSmiTak07}.

To construct $L$-operators we have to consider representations $\chi_\zeta$ and $\psi_\zeta$ of
$U_\hbar(\gothb'_+(A))$ and $U_\hbar(\gothb'_-(A))$, respectively, in addition to the
representation $\varphi_\zeta$ already described in section \ref{teh}. Slightly more abstractly, we
will use homomorphisms $\chi_\zeta$ and $\psi_\zeta$ of $U_\hbar(\gothb'_+(A))$ and
$U_\hbar(\gothb'_-(A))$ to $\rmOsc_\hbar$. It is easy to switch to representations, when necessary,
using the well known representations of $\rmOsc_\hbar$.

By an $L$-operator of type $\hat{L}$ we denote an element of $\rmOsc_\hbar \otimes \End(\bbC^3)
\simeq \mathrm{Mat}_3(\rmOsc_\hbar)$ defined as
\[
\hat{L}(\zeta_{12}) = \chi_{\zeta_1} \otimes \varphi_{\zeta_2} (\calR).
\]
Here, we have assumed that the homomorphisms $\chi_\zeta$ and $\varphi_\zeta$ are such that they
result in an $L$-operator depending only on $\zeta_{12} = \zeta_1/\zeta_2$. It follows from the
Yang--Baxter equation (\ref{ybe1}) that the $L$-operators of type $\hat{L}$ should satisfy the
equation
\begin{equation}
R_{23}(\zeta_{12}) \, \hat{L}_{13}(\zeta_1) \, \hat{L}_{12}(\zeta_2) = \hat{L}_{12}(\zeta_2) \,
\hat{L}_{13}(\zeta_1) \, R_{23}(\zeta_{12}). \label{rhlhl}
\end{equation}
As an $L$-operator of type $\check{L}$ we define an element of $\End(\bbC^3) \otimes \rmOsc_\hbar
\simeq \mathrm{Mat}_3(\rmOsc_\hbar)$
\[
\check{L}(\zeta_{12}) = \varphi_{\zeta_1} \otimes \psi_{\zeta_2} (\calR),
\]
assuming again that the homomorphisms $\psi_\zeta$ and $\varphi_\zeta$ are such that they result in
a dependence on the ratio $\zeta_{1}/\zeta_{2}$. Using the Yang--Baxter equation (\ref{ybe1}) we
derive the following equation for the $L$-operators of type $\check{L}$:
\begin{equation}
R_{12}(\zeta_{12}) \, \check{L}_{13}(\zeta_1) \, \check{L}_{23}(\zeta_2) = \check{L}_{23}(\zeta_2)
\, \check{L}_{13}(\zeta_1) \, R_{12}(\zeta_{12}). \label{rclcl}
\end{equation}
In terms of the matrices $\hat{R}$ and $\check{R}$ defined by means of the matrix of the
permutation operator $P_{12}$ as
\begin{equation}
\hat{R}(\zeta) = R(\zeta) P, \qquad \check{R}(\zeta) = P R(\zeta), \label{hRcR}
\end{equation}
equations (\ref{rhlhl}) and (\ref{rclcl}) take the forms
\[
\hat{R}(\zeta_{12}) ( \hat{L}(\zeta_1) \boxtimes \hat{L}(\zeta_2) ) = ( \hat{L}(\zeta_2) \boxtimes
\hat{L}(\zeta_1) ) \hat{R}(\zeta_{12})
\]
and
\[
\check{R}(\zeta_{12}) ( \check{L}(\zeta_1) \boxtimes \check{L}(\zeta_2) ) = ( \check{L}(\zeta_2)
\boxtimes \check{L}(\zeta_1) ) \check{R}(\zeta_{12}),
\]
respectively, where $\boxtimes$ means a generalization of the Kronecker product to the matrices with
arbitrary algebra-valued entries \cite{ChaPre95, BooGoeKluNirRaz10}.

Remember that the Borel subalgebra $U_\hbar(\gothb'_+(A))$ is generated by the elements
$h_i$, $e_i$, while the dual Borel subalgebra $U_\hbar(\gothb'_-(A))$ is generated by
$h_i$, $f_i$, $i = 0,1$. Here, the corresponding Serre relations (\ref{q-serre1}), (\ref{q-serre2})
are explicitly of the form
\begin{multline*}
{}\hspace{1.6cm} e_{\delta-2\alpha}^2 \, e_\alpha - (q^2 + q^{-2}) e_{\delta-2\alpha} \, e_\alpha
\, e_{\delta-2\alpha} + e_\alpha \, e_{\delta-2\alpha}^2 = 0, \\
e_\alpha^5 \, e_{\delta-2\alpha} - [5]_{q^{1/2}} e_\alpha^4 \, e_{\delta-2\alpha} \, e_\alpha +
\frac{[4]_{q^{1/2}} [5]_{q^{1/2}}}{[2]_{q^{1/2}}} (e_\alpha^3 \, e_{\delta-2\alpha} \, e_\alpha^2 -
e_\alpha^2 \, e_{\delta-2\alpha} \, e_\alpha^3) \\ + [5]_{q^{1/2}} e_\alpha \, e_{\delta-2\alpha}
\, e_\alpha^4 - e_{\delta-2\alpha} \, e_\alpha^5 = 0,
\end{multline*}
and we have the same equations for $f_{\delta-2\alpha}$ and $f_\alpha$. Besides, the Cartan
generators have to satisfy equations (\ref{qdr2}) and the condition $h_{\delta-2\alpha} + 2
h_{\alpha} = 0$.

To fulfill the defining relations of $U_\hbar(\gothb'_+(A))$, including also the
corresponding Serre relations, we use the homomorphism $\chi$ defined by the equations
\begin{align*}
& \chi(h_{\delta - 2\alpha}) = 2 D, \qquad\qquad\qquad\qquad \chi(h_{\alpha}) = -D, \\
& \chi(e_{\delta-2\alpha}) = {\mu_0} \, a^{\dagger \, 2} \, q^{-2(\nu + 1)D}, \qquad \chi(e_\alpha)
= {\mu_1} \, a \, q^{\nu D}
\end{align*}
with free parameters $\mu_0$, $\mu_1$ and $\nu$. The $\zeta$-dependent homomorphism $\chi_\zeta$
can be defined by the same procedure used earlier for the homomorphism $\varphi_\zeta$. Here,
changing the parameter $\mu_0$ we always change the coefficient at $\zeta^s$. Note that the
parameters $\mu_0$, $\mu_1$, $\nu$ can be freely changed by the transformations (\ref{2pag}).

Analogously, to satisfy the defining relations of $U_\hbar(\gothb'_-(A))$, meaning also the
corresponding Serre relations, we use the homomorphism $\psi$ defined by the equations
\begin{align*}
& \psi(h_{\delta - 2\alpha}) = - 2 D, \qquad\qquad\qquad\quad \psi(h_{\alpha}) = D, \\
& \psi(f_{\delta-2\alpha}) = {\mu_0} \, a^{\dagger \, 2} \, q^{-2(\nu + 1)D}, \qquad
\psi(f_\alpha) = \mu_1 \, a \, q^{\nu D},
\end{align*}
where $\mu_0$, $\mu_1$ and $\nu$ are again free parameters. The $\zeta$-dependent homomorphism
$\psi_\zeta$ can be defined by the same procedure used earlier for the homomorphism
$\varphi_\zeta$. Changing the parameter $\mu_0$ we always change the coefficient at $\zeta^s$, and
the parameters $\mu_0$, $\mu_1$, $\nu$ can be freely changed by the transformations (\ref{2pag}).
In both cases, the $L$-operators corresponding to different values of $\mu_0$, $\mu_1$ and $\nu$
are equivalent.

\subsection{\texorpdfstring{$L$-operators of type $\hat{L}$}{L-operators of type L}}
\label{thl}

Thus, to construct $L$-operators of type $\hat{L}$ we use the homomorphism $\chi_\zeta$ from
$U_\hbar(\gothb'_+(A))$ to $\rmOsc_\hbar$ defined by the following relations:\footnote{As above,
also in what follows the more convenient combination of the integers $s = s_0 + 2s_1$ and $s_1$
will be used instead of the initial ones, $s_0$ and $s_1$.}
\begin{align}
& \chi_\zeta(h_{\delta - 2\alpha}) = 2 D, \qquad\qquad\qquad\qquad\qquad\quad
\chi_\zeta(h_{\alpha}) = -D, \label{chi1} \\
& \chi_\zeta(e_{\delta-2\alpha}) = \frac{1}{(q-1)\sqrt{[2]_q}} \, a^{\dagger 2} \, q^{-2D} \,
\zeta^{s_0}, \qquad \chi_\zeta(e_{\alpha}) = \frac{1}{q-q^{-1}} \, a \, \zeta^{s_1}. \label{chi2}
\end{align}
This corresponds to the choice of the free parameters $\mu_0 = (q-1)^{-1}([2]_q)^{-1/2}$, $\mu_1 =
1/(q-q^{-1})$ and $\nu = 0$.

The higher root vectors are defined according to the recursive relations
(\ref{hivec1})--(\ref{hivec5}). Applying the homomorphism $\chi_\zeta$ we subsequently obtain
\[
\chi_\zeta (e'_{m\delta}) = \frac{q^m \zeta^{ms}}{(q-1)(q-q^{-1})} \, [(1 - q^{2m-1}) - (1 -
q^{2m+1})q^{2D}] \, q^{2(m-1)D},
\]
that gives
\[
\chi_\zeta (e_{m\delta}) = - \frac{q^m \zeta^{ms}}{m(q-q^{-1})} \, \left(1 - \frac{m}{[m]_q} \, b_m
\, q^m \, q^{2mD}\right),
\]
where the quantities $b_m$ are given explicitly by (\ref{bn}). Taking also the expression for
$\varphi_\zeta(f_{m\delta})$ from equation (\ref{fmd}) we obtain for the image of the factor
$\calR_{\sim \delta}$ the following expression:
\begin{multline*}
\chi_{\zeta_1} \otimes \varphi_{\zeta_2} (\calR_{\sim \delta}) = \rme^{{\lambda}(-q\zeta_{12}^s)}
\left( (1 + q^2 \zeta^s_{12} q^{2D}) \, E_{11} + \frac{\displaystyle (1 + q^2\zeta^s_{12})(1 - q^3
\zeta^s_{12} q^{2D})}{\displaystyle 1 + q^4 \zeta^s_{12} q^{2D}} \, E_{22} \right. \\
\left. + \frac{\displaystyle (1 + q^2 \zeta^s_{12})(1 - q^3 \zeta^s_{12})}{\displaystyle 1 - q^5
\zeta^s_{12} q^{2D}} \, E_{33} \right).
\end{multline*}

The simplest part of the calculations is, as usual, given by the operator $\calK$. In our case it
is represented by a diagonal matrix of the form
\[
\chi_{\zeta_1} \otimes \varphi_{\zeta_2} (\calK) = q^{-D} \, E_{11} + E_{22} + q^D \, E_{33}.
\]

As we already did for the $R$-matrix in section \ref{urm}, we can rearrange the factors entering
$\chi_{\zeta_1} \otimes \varphi_{\zeta_2} (\calR_{\prec \delta})$ in the definition of the
universal $R$-matrix in such a way that the factors corresponding to the roots $\alpha +m\delta$
come first, and only then come the factors corresponding to the roots $2\alpha + (2m+1)\delta$.
Similarly, concerning the factor $\chi_{\zeta_1} \otimes \varphi_{\zeta_2}(\calR_{\succ \delta})$,
we rearrange the factors entering there in such a way that the ones corresponding to the roots
$\delta - 2\alpha + 2m\delta$ come first, and finally come those ones corresponding to $\delta -
\alpha + m\delta$.

Altogether, the relations derived in Appendix \ref{clcLops} allow us to write down expressions corresponding to the factors
$\calR_{\prec \delta} = \calR_{\alpha} \calR_{2\alpha}$ and $\calR_{\succ \delta} = \calR_{\delta -
2\alpha} \calR_{\delta - \alpha}$. We obtain
\begin{multline*}
\chi_{\zeta_1} \otimes \varphi_{\zeta_2} (\calR_{\prec \delta}) = I + \zeta^{s_1}_{12} \, a
\, (1 + q^2 \, \zeta^s_{12} \, q^{2D})^{-1} \, E_{21} \\[.5em]
+ \zeta^{s_1}_{12} \, a \, (1 - q^3 \, \zeta^s_{12} \, q^{2D})^{-1} \, E_{32} + \frac{q}{q+1} \,
\zeta^{2s_1}_{12} \, a^2 \, (1 + q^2 \, \zeta^s_{12} \, q^{2D})^{-1} \, E_{31}
\end{multline*}
and
\begin{multline*}
\chi_{\zeta_1} \otimes \varphi_{\zeta_2} (\calR_{\succ \delta}) = I - \frac{q-q^{-1}}{q-1} \, q^2
\, \zeta^{s-s_1}_{12} \, a^\dagger \, (1 + q^4 \, \zeta^s_{12} \, q^{2D})^{-1} \,
E_{12} \\[.5em] + \frac{q-q^{-1}}{q-1} \, q^4 \, \zeta^{s-s_1}_{12} \, a^\dagger \,
(1 - q^5 \, \zeta^s_{12} \, q^{2D})^{-1} \, E_{23} \\[.5em]
+ \frac{q-q^{-1}}{q-1} \, \zeta^{s-2s_1}_{12} \, a^{\dagger 2} \, q^{-2D} \, (1 + q^6 \,
\zeta^{s}_{12} \, q^{2D})^{-1} \, E_{13}.
\end{multline*}
Multiplying the above presented factors in the given order, $\calR_{\prec \delta} \, \calR_{\sim
\delta} \, \calR_{\succ \delta} \, \calK$, we finally obtain the $L$-operator. According to our
terminology, the corresponding object is called an $L$-operator of type $\hat L$, and therefore, we
will use the respective notation. In matrix form we have
\[
\hat L(\zeta) = \rme^{\displaystyle \lambda(-q \zeta^s)} \left( \begin{array}{ccc} q^{-D} + q^2 \,
\zeta^s \, q^D & - (q+1)q \, \zeta^{s-s_1} \, a^\dagger & \frac{\displaystyle q+1}{\displaystyle q}
\, \zeta^{s-2s_1}
\, a^{\dagger 2} \, q^{-D} \\
\zeta^{s_1} \, a \, q^{-D} & 1 - q \, \zeta^s & \frac{\displaystyle q+1}{\displaystyle q} \,
\zeta^{s-s_1}
\, a^\dagger \, q^{-D} \\
\frac{\displaystyle q}{\displaystyle q+1} \, \zeta^{2s_1} \, a^2 \, q^{-D} & \zeta^{s_1} \, a & q^D
+ \zeta^s \, q^{-D}
\end{array} \right).
\]
We have the following relation between the $L$-operators with arbitrary values of the parameters
and fixed ones:
\[
\hat{L}^{(s,s_1)}(\zeta_{12}) = \gamma_{\zeta_1} \left( G(\zeta_2) \, \hat{L}^{(1,0)}(\zeta^s_{12})
\, G^{-1}(\zeta_2)\right),
\]
where the matrix $G(\zeta)$ was defined earlier by (\ref{Gz}), while the mapping $\gamma_\zeta$,
$\zeta \in \bbC^\times$, is defined by the relations
\[
\gamma_\zeta(a) = a \, \zeta^{-s_1}, \qquad \gamma_\zeta(a^\dagger) = a^\dagger \, \zeta^{s_1},
\qquad \gamma_\zeta(D) = D.
\]
It means, in particular, that the $Q$-operators based on $L$-operators with different values of $s$
and $s_1$ are related by a change of the spectral parameters and a similarity transformation.

Applying to $\hat{L}$ the automorphism $\sigma$ generated by the transformation
\[
a^\dagger \to a \, q^{-D}, \qquad a \to - q^{-D} \, a^\dagger, \qquad D \to - D - 1,
\]
we obtain another $L$-operator of type $\hat{L}$,
\[
\sigma : \hat L \to \hat L_\sigma,
\]
where
\[
\hat L_\sigma(\zeta) =  \rme^{\displaystyle \lambda(-q\zeta^s)} \, \hat L'_\sigma(\zeta),
\]
and we have used the notation
\[
\hat L'_\sigma (\zeta) = \left( \begin{array}{ccc} q(q^{D} + \, \zeta^s \, q^{-D})
& - (q+1)q \, \zeta^{s-s_1} \, a \, q^{-D} & (q+1)q \, \zeta^{s-2s_1} \, a^2 \, q^{-D} \\
- \zeta^{s_1} \, a^\dagger & 1 - q \, \zeta^s & (q+1) \, \zeta^{s-s_1} \, a \\
\frac{\displaystyle q^{-1}}{\displaystyle q+1} \, \zeta^{2s_1} \, a^{\dagger 2} \, q^{-D} & -
q^{-1} \zeta^{s_1} \, a^\dagger \, q^{-D} & q^{-1}(q^{-D} + q^2 \, \zeta^s \, q^{D})
\end{array} \right).
\]

Note also that an object similar to $\hat{L}'_\sigma$ was used in \cite{CorZam10} in attempts to
describe so-called defects in affine Toda field theory.

\subsection{\texorpdfstring{Type ${\check{L}}$ and some useful relations}{Type L and some useful relations}}
\label{tcl}

The $L$-operators of type $\check L$ can be constructed by means of the homomorphisms
$\varphi_\zeta$ and $\psi_\zeta$. This requires to map the Borel subalgebra
$U_\hbar(\gothb'_+(A))$ to the finite-dimensional matrix representation, while realizing
the Borel subalgebra $U_\hbar(\gothb'_-(A))$ in $\rmOsc_\hbar$. Here we consider the
following homomorphism for the generators $f_{\alpha_0}$, $f_{\alpha_1}$ and $h_{\alpha_0}$,
$h_{\alpha_1}$:
\[
\psi_{\zeta} (f_{\delta-2\alpha}) = \frac{1}{(q-1)\sqrt{[2]_q}} \, a^{\dagger 2} \, q^{-2D} \,
\zeta^{-s_0}, \qquad \psi_\zeta (f_{\alpha}) = \frac{1}{q-q^{-1}} \, a \, \zeta^{-s_1}
\]
and
\[
\psi_\zeta (h_{\delta - 2\alpha}) = -2D, \qquad \psi_\zeta (h_{\alpha}) = D,
\]
Using the Cartan anti-involution we obtain from the recursive relations
(\ref{hivec1})--(\ref{hivec5}) recursive relations for the higher root vectors spanning the whole
$U_\hbar(\gothb'_-(A))$. Performing the whole procedure in the same way as in the preceding
subsection, we obtain the following matrix form of the $L$-operator of type $\check{L}$:
\[
\check L(\zeta) = \rme^{\displaystyle \lambda(-q\zeta^s)} \left( \begin{array}{ccc} q^{D} + \zeta^s
\, q^{-D} & \zeta^{s_1} \, a &
\frac{\displaystyle q}{\displaystyle q+1} \, \zeta^{2s_1} \, a^{2} \, q^{-D} \\
\frac{\displaystyle q+1}{\displaystyle q} \, \zeta^{s-s_1} \, a^\dagger \, q^{-D} & 1 - q \,
\zeta^s & \zeta^{s_1} \, a \, q^{-D} \\
\frac{\displaystyle q+1}{\displaystyle q}\zeta^{s-2s_1} \, a^{\dagger 2} \, q^{-D} & - (q+1)q \,
\zeta^{s-s_1} \, a^\dagger & q^{-D} + q^2 \, \zeta^s \, q^{D}
\end{array} \right).
\]
Here, to bring the scalar factor to the given simple form, we have used the relation
\begin{equation}
\frac{\displaystyle \rme^{\displaystyle - \lambda(q^4\zeta)}}{\displaystyle (1 + q^2 \zeta)(1 -
q^3 \zeta)} = \rme^{\displaystyle \lambda(-q\zeta)} \label{ssf}
\end{equation}
following from the basic equation (\ref{lbr}) for the function $\lambda$. For the $L$-operators of
type $\check{L}$ we have the following relation:
\[
\check{L}^{(s,s_1)}(\zeta_{12}) = G(\zeta_1) \, \gamma_{\zeta_2} \left(
\check{L}^{(1,0)}(\zeta^s_{12}) \right) G^{-1}(\zeta_1).
\]

Now, comparing the explicit forms of the $L$-operators of the type $\hat L$ and $\check L$ we see
that they are related as
\[
\check L(\zeta) = J \, \hat L(\zeta) \, J,
\]
where $J$ is the $3 \times 3$ skew-diagonal unit matrix.\footnote{This expression $J x J$ implies
actually the superposition of two operations, where one has to take first the usual transposition
of $x$ with respect to its main diagonal, and then the transposition of the obtained matrix with
respect to the skew diagonal.} Besides, for the matrices $\hat{R}$ and $\check{R}$ introduced by
equation (\ref{hRcR}), we obtain the following relation:
\[
\check R = (J \otimes J) \, \hat R \, (J \otimes J).
\]

Also the inverse of the $L$-operator of type $\hat L$ at $\zeta^{-1}$ should be an $L$-operator of
type $\check L$. Explicitly we have
\[
\hat{L}^{-1}(\zeta) = \frac{\displaystyle \rme^{\displaystyle - \lambda(-q\zeta^s)}}{\displaystyle
(1 + q^2 \zeta^s)(1 - q^3 \zeta^s)} \, \check{L}'(\zeta),
\]
where we have denoted
\[
\check L'(\zeta) = \left( \begin{array}{ccc} q^{D} - q^3 \, \zeta^s \, q^{-D} & (q+1)q^3 \,
\zeta^{s-s_1} \, q^{-D} \, a^\dagger & - (q+1)q^3 \, \zeta^{s-2s_1}
\, q^{-D} \, a^{\dagger 2} \\
- \zeta^{s_1} \, a & 1 + q^4 \, \zeta^s & -(q+1)q^3 \, \zeta^{s-s_1}
\, a^\dagger \\
\frac{\displaystyle 1}{\displaystyle q+1} \, \zeta^{2s_1} \, q^{-D} \, a^2 & - \zeta^{s_1} \,
q^{-D} \, a & q^{-D} - q^5 \, \zeta^s \, q^{D}
\end{array} \right).
\]
And again, using equation (\ref{ssf}) one can simplify the form of the scalar factor at
$\check{L}'$.

Note finally that, if $\hat{L}(\zeta)$ is an $L$-operator of type $\hat L$, then
$\tau(\hat{L}(\zeta^{-1}))$ is an $L$-operator of type $\check{L}$, and vice versa, if
$\check{L}(\zeta)$ is an $L$-operator of type $\check L$, then $\tau(\check{L}(\zeta^{-1}))$ is an
$L$-operator of type $\hat{L}$, where $\tau$ means the anti-involution of the $q$-oscillator
algebra $\rmOsc_\hbar$ defined by the relations
\[
\tau(a) = a^\dagger, \qquad \tau(a^\dagger) = a, \qquad \tau(D) = D.
\]
Besides, applying to $\check{L}$ the automorphism $\sigma$ described in the preceding sub-section,
one can obtain another $L$-operator of type $\check{L}$.

\section{Concluding remarks}
\label{cor}

We have constructed the $R$-matrix for the twisted Kac--Moody algebra of type $A^{(2)}_2$, that is
given by the universal $R$-matrix under the action of a twisted evaluation homomorphism. That
latter allowed us to bring the dual Borel subalgebras to a finite-dimensional matrix
representation. We have found an expression for the respective spin-chain Hamiltonian. We have also
constructed $L$-operators realizing one of the Borel subalgebras in the $q$-deformed oscillator
algebra.

We have seen, in particular, that the $R$-matrix and $L$-operators have two points of degeneracy in
this twisted case in contrast to the earlier considered untwisted cases where we had only one such
a point.

Recall that one usually considers a useful decomposition of the $R$-matrix in the form
\[
R(z) = z R_0 - z^{-1} R_0^{-1},
\]
where the non-degenerate matrix $R_0$ does not depend on the spectral parameter $z$. Instead,
taking, for simplicity $s_0 = 1$, $s_1 = 0$, so that $s = 1$, we will have another relation,
\[
R(\zeta) = \Lambda(\zeta) \left[ q^{1/2} ( \zeta^{-1} \, r_1 - r_0 ) - \Omega_9 \left( q^{1/2} (
\zeta^{-1} \, r_1 - r_0 ) \right) \right],
\]
where the upper-triangular matrices $r_0$ and $r_1$ do not depend on
$\zeta$ and the scalar pre-factor $\Lambda$ is a rational function
subject to the condition
\[
\Omega_1(\Lambda(\zeta)) = -q^{-2} \Lambda(\zeta).
\]
Here we explicitly have
\begin{multline*}
r_0 = E_{11} \otimes E_{11} + q \, E_{11} \otimes E_{22} + E_{11}
\otimes E_{33} \\ + q \left( E_{22} \otimes E_{11} + (1 -
(q-q^{-1})) E_{22} \otimes E_{22} + E_{22} \otimes E_{33} \right) \\
+ E_{33} \otimes E_{11} + q \, E_{33} \otimes E_{22} + E_{33}
\otimes E_{33} \\ - (q-q^{-1}) \left( q^{-2} \, E_{12} \otimes
E_{21} - q \, E_{12} \otimes E_{32} - E_{23} \otimes E_{21} \right.
\\ \left. + (q + q^{-1}) q \, E_{13} \otimes E_{31} + q^{-2} \, E_{23}
\otimes E_{32} \right),
\end{multline*}
\begin{multline*}
r_1 = q^2 \, E_{11} \otimes E_{11} + q \, E_{11} \otimes E_{22} +
E_{11} \otimes E_{33} \\ + q \left( E_{22} \otimes E_{11} + E_{22} \otimes E_{22}
+ E_{22} \otimes E_{33} \right) \\
+ E_{33} \otimes E_{11} + q \, E_{33} \otimes E_{22} + q^2 \, E_{33}
\otimes E_{33} \\ + (q-q^{-1}) \left( q \, E_{12} \otimes E_{21} + q
\, E_{12} \otimes E_{32} + E_{23} \otimes E_{21} \right.
\\ \left. + (q - 1) \, E_{13} \otimes E_{31} + q \, E_{23}
\otimes E_{32} \right)
\end{multline*}
and
\[
\Lambda(\zeta) = \frac{q^{-3/2} \zeta}{(1-q^{-2}\zeta)(1+q^{-3}\zeta)}.
\]

We have also constructed $L$-operators of two types, $\hat{L}$ and $\check{L}$, in the
$q$-oscillator representation. They satisfy Yang--Baxter equations with the $R$-matrices $\hat{R}$
and $\check{R}$, respectively. Here, the $L$-operators and $R$-matrices of different types are
related by similarity transformations.

The $L$-operators allow for the decomposition of the usual form $\zeta L_+ - \zeta^{-1} L_-$, where
the constituents $L_+$ and $L_-$ are presented by non-degenerate matrices. The latter differs from
the untwisted case, where one of such two parts, $L_+$ or $L_-$, turned out to be degenerate
\cite{BooGoeKluNirRaz10}. However, to find also for the $L$-operators a decomposition similar to
what we have for the $R$-matrix, one needs to define an analogue of the Cartan anti-involution in
the $q$-oscillator algebra.

An interesting and quite non-trivial question in a lattice model is about continuum analogues of any monodromy type matrices. In a reasonable classical limit one should anticipate that the corresponding objects, in terms of suitable phase space variables, would have certain fundamental Poisson brackets with the classical $r$-matrix. This problem, in a somewhat reversed form, was formulated in the pioneering papers on the quantum inverse scattering method, see, for example, \cite{SklTakFad79, IzeKor82}. And also in our case it will be the matter of further investigation.

Another special question concerns the quasi-classical limit. If we evaluate our $L$-operators naively at $\hbar \to 0$, where $q \to 1$ and $\zeta \to 1$, we obtain a rather meaningless degenerate matrix. Instead, we could follow an idea exploited e.~g. in \cite{BazLukMenSta11} and first renormalize the $L$-operators by means of certain linear transformation. Let us consider $\hat{L}(\zeta)$ of section \ref{thl} with fixed $s=2$, $s_1=0$ and change there $\zeta^2$ by $-\zeta^2$ for convenience. Performing now the transformation
\[
\hat{L}(\zeta) \to \zeta^{-1} \, \rme^{\displaystyle -\lambda(q\zeta^2)} \, \calI_q \, \hat{L}(\zeta) \, {}^{J\!}\calI_q,
\]
where
\[
\calI_q = \left( \begin{array}{ccc} 1 & & \\
                                        & 1 & \\
                                        & & - 1/(q-q^{-1}) \end{array} \right), \qquad
{}^J\calI_q = J \, \calI_q \, J,
\]
we rewrite the transformed $L$-operator in terms of the naturally $q$-deformed oscillators\footnote{See the footnote in section \ref{rsr}.} $\calN$, $b$ and $b^\dagger$ and use the parametrization $q = \rme^\hbar$ and $\zeta = \rme^{\hbar z}$. Then, in the limit $\hbar \to 0$ we have
\[
\hat{L}_c(z) = \left( \begin{array}{ccc}
                         z + \calN + 1 & 2 b^\dagger & -2 b^{\dagger 2} \\
                         b & 2 & -2 b^\dagger \\
                         b^2/2 & b & z - \calN \end{array} \right).
\]
One can treat the obtained matrix $\hat{L}_c(z)$ as a local quantum Lax operator for the quasi-classical limit of the Izergin--Korepin model.

The objects constructed in this work will be used in future work to develop the method of
functional relations including the transfer matrix and Baxter's $Q$-operators for the quantum
integrable system considered here. Note that to obtain functional relations one can also use an
approach based on the notion of fundamental modules over the Borel subalgebras of the quantum
algebras \cite{HerJim11, JimSun10}, allowing one to avoid explicit forms of the $L$-operators.

\vskip5mm {\em Acknowledgements.\/} We are grateful to M. Jimbo for communication. This work was
supported in part by the Volkswagen Foundation. A.V.R. was supported in part by the RFBR grants \#
09-01-93107 and \# 10-01-00300.

\appendix

\section{Calculating the $R$-matrix}
\label{clcRmat}

The calculation of the factors $\varphi_{\zeta_1} \otimes \varphi_{\zeta_2} (\calR_{2\alpha})$ and
$\varphi_{\zeta_1} \otimes \varphi_{\zeta_2} (\calR_{\delta-2\alpha})$ belongs obviously to the
simplest part of the work, since, due to the relation
\[
(E_{ij})^\ell = 0, \qquad i \neq j, \qquad \ell > 1,
\]
the infinite products in the corresponding expressions reduce to simple infinite geometric series.
Indeed, we have
\begin{multline*}
\varphi_{\zeta_1} \otimes \varphi_{\zeta_2} (\calR_{2\alpha}) = \prod^\curvearrowright_{m \ge 0}
\left( I + (q-q^{-1}) \, \varphi_{\zeta_1} (e_{2\alpha +
(2m+1)\delta}) \otimes \varphi_{\zeta_2} (f_{2\alpha + (2m+1)\delta}) \right) \\
= I + (q-q^{-1}) \sum_{m \ge 0} \varphi_{\zeta_1}(e_{2\alpha +
(2m+1)\delta}) \otimes \varphi_{\zeta_2} (f_{2\alpha + (2m+1)\delta}),
\end{multline*}
where $I$ stands for the $9 \times 9$ unit matrix. Then we easily perform the summation in the expression
above and obtain
\[
\varphi_{\zeta_1} \otimes \varphi_{\zeta_2} (\calR_{2\alpha}) = I + (q-q^{-1}) \, [2]_q \,
\frac{\zeta^{s + 2s_1}_{12}}{1 - \zeta^{2s}_{12}} \, E_{13} \otimes E_{31}.
\]

Similarly, for the factor $\calR_{\delta-2\alpha}$ we obtain
\begin{multline*}
\varphi_{\zeta_1} \otimes \varphi_{\zeta_2} (\calR_{\delta-2\alpha}) = \prod^\curvearrowleft_{m \ge
0} \left( I + (q-q^{-1}) \, \varphi_{\zeta_1} (e_{\delta -
2\alpha + 2m\delta})
\otimes \varphi_{\zeta_2} (f_{\delta - 2\alpha + 2m\delta}) \right) \\
= I + (q-q^{-1}) \sum_{m \ge 0} \varphi_{\zeta_1}
(e_{\delta-2\alpha + 2m\delta}) \otimes \varphi_{\zeta_2} (f_{\delta - 2\alpha + 2m\delta}).
\end{multline*}
This allows us to write down the expression
\[
\varphi_{\zeta_1} \otimes \varphi_{\zeta_2} (\calR_{\delta - 2\alpha}) = I + (q-q^{-1}) \, [2]_q \,
\frac{\zeta^{s-2s_1}_{12}}{1 - \zeta^{2s}_{12}} \, E_{31} \otimes E_{13}.
\]

The most complicated part of our calculations concerns the factors $\calR_{\alpha}$ and
$\calR_{\delta - \alpha}$. This is due to the fact that the images of the generators in the
corresponding exponentials are now nilpotent of degree three. Hence, we need to take into account
terms quadratic in such generators,
\[
\varphi_{\zeta_1} \otimes \varphi_{\zeta_2} (\calR_{\alpha}) = \prod^{\curvearrowright}_{m \ge 0}
\exp_{q_\alpha} (x_{\alpha,m}) = \prod^\curvearrowright_{m \ge 0} \left( 1 + x_{\alpha,m} +
\frac{x^2_{\alpha,m}}{(2)_{q_\alpha}!} \right),
\]
while all higher order terms vanish. Here, for convenience, we have used the notation
\[
x_{\alpha,m} = (q-q^{-1}) \, \varphi_{\zeta_1} (e_{\alpha + m\delta})
\otimes \varphi_{\zeta_2} (f_{\alpha + m\delta}),
\]
where $\varphi_{\zeta_1} (e_{\alpha+m\delta})$ and $\varphi_{\zeta_2} (f_{\alpha+m\delta})$ are
explicitly given above. The infinite product in the expression for $\varphi_{\zeta_1}
\otimes \varphi_{\zeta_2} (\calR_{\alpha})$ can be rewritten as the following infinite sum:
\[
\varphi_{\zeta_1} \otimes \varphi_{\zeta_2} (\calR_\alpha) = I + \sum_{m \ge 0} x_{\alpha,m} +
\frac{q}{q+1} \sum_{m \ge 0} x^2_{\alpha,m} + \sum_{m \ge 0} \sum_{k \ge 0} x_{\alpha,m} \,
x_{\alpha,k+m+1}.
\]
Performing the corresponding summations,
\begin{align*}
\sum_{m \ge 0} x_{\alpha,m} & =  \frac{q-q^{-1}}{1-\zeta^s_{12}} \, \zeta^{s_1}_{12} \, \left(
E_{12} \otimes E_{21} + E_{23} \otimes E_{32} \right) \\
& \hskip7mm + \frac{q-q^{-1}}{1+q\zeta^s_{12}} \, \zeta^{s_1}_{12} \, E_{12} \otimes E_{32} +
\frac{q-q^{-1}}{1+q^{-1}\zeta^s_{12}} \, \zeta^{s_1}_{12} \, E_{23} \otimes E_{21}, \\[.3em]
\frac{q}{q+1} \sum_{m \ge 0} x^2_{\alpha,m} & =
\frac{(q-q^{-1})(q-1)}{1-\zeta^{2s}_{12}} \, \zeta^{2s_1}_{12} \, E_{13} \otimes E_{31}, \\[.3em]
\sum_{m \ge 0} x_{\alpha,m} \sum_{k \ge 0} x_{\alpha,k+m+1} & = - \frac{(q - q^{-1})^2
q^{-1}\zeta^{s+2s_1}_{12}}{(1+q^{-1}\zeta^s_{12})(1-\zeta^{2s}_{12})} \, E_{13} \otimes E_{31},
\end{align*}
we obtain
\begin{multline*}
\varphi_{\zeta_1} \otimes \varphi_{\zeta_2} (\calR_{\alpha}) = I + \frac{q-q^{-1}}{1-\zeta^s_{12}}
\, \zeta^{s_1}_{12} \left( E_{12} \otimes E_{21} + E_{23} \otimes E_{32} \right) \\
+ \frac{q-q^{-1}}{1+q\zeta^s_{12}} \, \zeta^{s_1}_{12} \, E_{12} \otimes E_{32} +
\frac{q-q^{-1}}{1+q^{-1}\zeta^s_{12}} \, \zeta^{s_1}_{12} \, E_{23} \otimes E_{21} \\
+ \frac{(q-q^{-1})(q-1)(1-q^{-2}\zeta^s_{12})}{(1-\zeta^{2s}_{12})(1+q^{-1}\zeta^s_{12})} \,
\zeta^{2s_1}_{12} \, E_{13} \otimes E_{31}.
\end{multline*}

We have the same situation with the factor $\calR_{\delta-\alpha}$. Here we have to carry out the
expression
\[
\varphi_{\zeta_1} \otimes \varphi_{\zeta_2} (\calR_{\delta-\alpha}) = \prod^{\curvearrowleft}_{m
\ge 0} \exp_{q_\alpha} (y_{\alpha,m}) = \prod^\curvearrowleft_{m \ge 0} \left( 1 + y_{\alpha,m} +
\frac{y^2_{\alpha,m}}{(2)_{q_\alpha}!} \right),
\]
where now we have denoted
\[
y_{\alpha,m} = (q-q^{-1}) \, \varphi_{\zeta_1} (e_{\delta -
\alpha + m\delta}) \otimes \varphi_{\zeta_2} (f_{\delta - \alpha + m\delta}),
\]
with $\varphi_{\zeta_1} (e_{\delta - \alpha + m\delta})$ and $\varphi_{\zeta_2} (f_{\delta - \alpha
+ m\delta})$ given in the preceding section. Again, we rewrite the infinite product in the expression
for $\varphi_{\zeta_1} \otimes
\varphi_{\zeta_2} (\calR_{\delta - \alpha})$ as an infinite sum,
\[
\varphi_{\zeta_1} \otimes \varphi_{\zeta_2} (\calR_{\delta - \alpha}) = I + \sum_{m \ge 0}
y_{\alpha,m} + \frac{q}{q+1} \sum_{m \ge 0} y^2_{\alpha,m} + \sum_{m \ge 0} \sum_{k \ge 0}
y_{\alpha,k+m+1} \, y_{\alpha,m}.
\]
After some calculations,
\begin{align*}
\sum_{m \ge 0} y_{\alpha,m} & = \frac{(q-q^{-1})\zeta^{s-s_1}_{12}}{1-\zeta^s} \left(
E_{21} \otimes E_{12} + E_{32} \otimes E_{23} \right) \\
& \hskip7mm - \frac{(q-q^{-1})q^2\zeta^{s-s_1}_{12}}{1+q\zeta^s_{12}} \, E_{21} \otimes E_{23} -
\frac{(q-q^{-1})q^{-2}\zeta^{s-s_1}_{12}}{1+q^{-1}\zeta^s_{12}} \, E_{32} \otimes E_{12}, \\[.3em]
\frac{q}{q+1} \sum_{m \ge 0} y^2_{\alpha,m} & =
\frac{(q-q^{-1})(q-1)\zeta^{2(s-s_1)}_{12}}{1-\zeta^{2s}_{12}} \, E_{31} \otimes E_{13},
\\[.3em]
\sum_{m \ge 0} \sum_{k \ge 0} y_{\alpha,k+m+1} \, y_{\alpha,m} & = - \frac{(q - q^{-1})^2
q^{-1}\zeta^{3s-2s_1}_{12}}{(1+q^{-1}\zeta^s_{12})(1-\zeta^{2s}_{12})} \, E_{31} \otimes E_{13},
\end{align*}
we obtain
\begin{multline*}
\varphi_{\zeta_1} \otimes \varphi_{\zeta_2} (\calR_{\delta - \alpha}) = I +
\frac{(q-q^{-1})\zeta^{s-s_1}_{12}}{1-\zeta^s_{12}} \left(
E_{21} \otimes E_{12} + E_{32} \otimes E_{23} \right) \\
- \frac{(q-q^{-1})q^2\zeta^{s-s_1}_{12}}{1+q\zeta^s_{12}} \, E_{21} \otimes E_{23} -
\frac{(q-q^{-1})q^{-2}\zeta^{s-s_1}_{12}}{1+q^{-1}\zeta^s_{12}} \, E_{32} \otimes E_{12} \\
+ \frac{(q-q^{-1})(q-1-(1-q^{-1})q^{-1}\zeta^s_{12})\zeta^{2(s-s_1)}_{12}}
{(1-\zeta^{2s}_{12})(1+q^{-1}\zeta^s_{12})} \, E_{31} \otimes E_{13}.
\end{multline*}

\section{Calculating the $L$-operators}
\label{clcLops}

With respect to the higher roots $2\alpha +(2m+1)\delta$ we obtain the expression
\[
\chi_\zeta (e_{2\alpha+(2m+1)\delta})
= \frac{(q-1) [2]_q^{1/2} \zeta^{2s_1 + (2m+1)s}}{(q-q^{-1})(q^2-q^{-2})}
\, a^2 \, q^{2(2m+1)D},
\]
which, in addition to (\ref{hivec6a}) for $\varphi_\zeta(f_{2\alpha+(2m+1)\delta})$, leads to
\[
\chi_{\zeta_1} \otimes \varphi_{\zeta_2} (\calR_{2\alpha}) = I + \frac{q(q-1)}{q-q^{-1}} \,
\zeta^{s+2s_1}_{12} \, a^2 \, q^{2D} (1 + q^3 \, \zeta^{2s}_{12} \, q^{4D})^{-1} \, E_{31}.
\]
Related with the roots $\delta-2\alpha+2m\delta$ we obtain the expression
\[
\chi_\zeta (e_{\delta-2\alpha+2m\delta}) = \frac{q^{8m}\zeta^{s-2s_1 + 2ms}}{(q-1)[2]_q^{1/2}} \,
a^{\dagger 2} \, q^{2(2m-1)D},
\]
and so, using also equation (\ref{hivec8a}) for $\varphi_\zeta(f_{\delta-2\alpha+2m\delta})$, we
obtain
\[
\chi_{\zeta_1} \otimes \varphi_{\zeta_2} (\calR_{\delta-2\alpha}) = I + \frac{q-q^{-1}}{q-1} \,
\zeta^{s-2s_1}_{12} \, a^{\dagger 2} \, q^{-2D} (1 + q^{11} \, \zeta^{2s}_{12} \, q^{4D})^{-1} \,
E_{13}.
\]

Further, for the factor $\calR_\alpha$ we obtain the equation
\[
\chi_{\zeta_1} \otimes \varphi_{\zeta_2} (\calR_\alpha) = I + \sum_{m \ge 0} x_{\alpha,m} +
\frac{q}{q+1} \sum_{m \ge 0} x^2_{\alpha,m} + \sum_{m \ge 0} \sum_{k \ge 0} x_{\alpha,m} \,
x_{\alpha,k+m+1},
\]
where now
\[
x_{\alpha,m} = (q-q^{-1}) \, \chi_{\zeta_1} (e_{\alpha + m\delta}) \otimes
\varphi_{\zeta_2} (f_{\alpha + m\delta}).
\]
Using the expressions
\[
\chi_{\zeta} (e_{\alpha+m\delta}) = \frac{q^m \zeta^{s_1+ms}}{q-q^{-1}} \, a \, q^{2mD},
\]
and (\ref{hivec2a}) for $\varphi_\zeta(f_{\alpha+m\delta})$, we obtain
\begin{align*}
& \sum_{m \ge 0} x_{\alpha,m} = \zeta^{s_1}_{12} \, a \, [(1 + q^2 \, \zeta^s_{12} \, q^{2D})^{-1}
\, E_{21} + (1 - q^3 \, \zeta^s_{12} \, q^{2D})^{-1} \, E_{32}], \\[.5em]
& \frac{q}{q+1} \sum_{m \ge 0} x^2_{\alpha,m} = \frac{q\zeta^{2s_1}_{12}}{q+1} \, a^2 \, (1 + q^3
\, \zeta^{2s}_{12} \, q^{4D})^{-1} \, E_{31}, \\[.5em]
& \sum_{m \ge 0} \sum_{k \ge 0} x_{\alpha,m} \, x_{\alpha,k+m+1} = - q^2 \, \zeta^{s+2s_1}_{12} \,
a^2 \, q^{2D} \, (1 + q^3 \, \zeta^{2s}_{12} \, q^{4D})^{-1} (1 + q^2 \, \zeta^s_{12} \,
q^{2D})^{-1} \, E_{31}.
\end{align*}
This leads us to the expression
\begin{multline*}
\chi_{\zeta_1} \otimes \varphi_{\zeta_2} (\calR_\alpha)
= I + \zeta^{s_1}_{12} \, a \, (1 + q^2 \, \zeta^s_{12} \, q^{2D})^{-1} \, E_{21} \\
+ \zeta^{s_1}_{12} \, a \, (1 - q^3 \, \zeta^s_{12} \, q^{2D})^{-1} \, E_{32} + \frac{q}{q+1} \,
\zeta^{2s_1}_{12} \, a^2 \, (1 + q^3 \, \zeta^{2s}_{12} \, q^{4D})^{-1} \, E_{31} \\
- q^2 \, \zeta^{s+2s_1}_{12} \, a^2 \, q^{2D} \, (1 + q^3 \, \zeta^{2s}_{12} \, q^{4D})^{-1} (1 +
q^2 \, \zeta^s_{12} \, q^{2D})^{-1} \, E_{31}.
\end{multline*}

For the factor $\calR_{\delta-\alpha}$ we have the equation
\[
\chi_{\zeta_1} \otimes \varphi_{\zeta_2} (\calR_{\delta-\alpha}) = I + \sum_{m \ge 0} y_{\alpha,m}
+ \frac{q}{q+1} \sum_{m \ge 0} y^2_{\alpha,m} + \sum_{m \ge 0} \sum_{k \ge 0} y_{\alpha,k+m+1} \,
y_{\alpha,m},
\]
where we have used the notation
\[
y_{\alpha,m} = (q-q^{-1}) \, \chi_{\zeta_1} (e_{\delta - \alpha +
m\delta}) \otimes \varphi_{\zeta_2} (f_{\delta - \alpha + m\delta}).
\]
Using the expressions
\[
\chi_\zeta (e_{\delta-\alpha+m\delta}) = - \frac{q^{3m+2}\zeta^{s-s_1+ms}}{q-1} \, a^\dagger \,
q^{2mD}
\]
and (\ref{hivec4a}) for $\varphi_\zeta (f_{\delta-\alpha+m\delta})$, we obtain
\begin{align*}
\sum_{m \ge 0} y_{\alpha,m} & = - \frac{q-q^{-1}}{q-1} \, q^2 \, \zeta^{s-s_1}_{12} \, a^\dagger \,
[(1 + q^4 \, \zeta^s_{12} \, q^{2D})^{-1} \, E_{12} - q^2 \, (1 - q^5 \, \zeta^s_{12} \,
q^{2D})^{-1} \, E_{23}], \\[.5em]
\frac{q}{q+1} \sum_{m \ge 0} y^2_{\alpha,m} & = - \frac{q^7}{q+1} \frac{(q-q^{-1})^2}{(q-1)^2} \,
\zeta^{2(s-s_1)}_{12} \, a^{\dagger 2} \, (1 + q^{11} \, \zeta^{2s}_{12} \, q^{4D})^{-1} \, E_{13}, \\[.5em]
& \hspace{-2.5cm} \sum_{m \ge 0} \sum_{k \ge 0} y_{\alpha,k+m+1} \, y_{\alpha,m} \\[.5em]
& = \frac{(q-q^{-1})^2}{(q-1)^2} \, q^{12} \, \zeta^{3s-2s_1}_{12} \, a^{\dagger 2} \, q^{2D} \, (1
+ q^{11} \, \zeta^{2s}_{12} \, q^{4D})^{-1} (1 + q^6 \, \zeta^s_{12} \, q^{2D})^{-1} \, E_{13}.
\end{align*}
Summing up these terms, we obtain
\begin{multline*}
\chi_{\zeta_1} \otimes \varphi_{\zeta_2} (\calR_{\delta-\alpha}) = I - \frac{q-q^{-1}}{q-1} \, q^2
\, \zeta^{s-s_1}_{12} \, a^\dagger \, (1 + q^4 \, \zeta^s_{12} \, q^{2D})^{-1} \, E_{12} \\
+ \frac{q-q^{-1}}{q-1} \, q^4 \, \zeta^{s-s_1}_{12} \, a^\dagger \, (1 - q^5 \, \zeta^s_{12} \,
q^{2D})^{-1} \, E_{23} \\ + \frac{q^7}{q+1} \frac{(q-q^{-1})^2}{(q-1)^2} \, \zeta^{2(s-s_1)}_{12}
\, a^{\dagger 2} \, (1 + q^{11} \, \zeta^{2s}_{12} \, q^{4D})^{-1} \, E_{13} \\
+ \frac{(q-q^{-1})^2}{(q-1)^2} q^{12} \, \zeta^{3s-2s_1}_{12} \, a^{\dagger 2} \, q^{2D} \, (1 +
q^{11} \, \zeta^{2s}_{12} \, q^{4D})^{-1} (1 + q^6 \, \zeta^s_{12} \, q^{2D})^{-1} \, E_{13}.
\end{multline*}

\bibliographystyle{ihepengplain}

\bibliography{IntegrableSystems}

\newcommand{\noopsort}[1]{}
\providecommand{\bysame}{\leavevmode\hbox to3em{\hrulefill}\thinspace}
\providecommand{\href}[2]{{#2}}

\end{document}